\documentclass[graybox, envcountchap]{svmult}

\usepackage{mathptmx}        
\usepackage{amsmath}
\usepackage{amssymb}
\usepackage{color}
\usepackage{helvet}          
\usepackage{courier}         
\usepackage{dirtree}

\usepackage{makeidx}        
\usepackage{graphicx}        
\usepackage{subfig}

\usepackage{multicol}        
\usepackage[bottom]{footmisc}

\usepackage{hyperref}        
\hypersetup{colorlinks=true,urlcolor=blue}

\usepackage[misc]{ifsym}

\makeindex             

\begin{document}


\title{Alternative Derivations of Hawking Radiation}
\author{Chon Man Sou}
\institute{Chon Man Sou (\Letter) \at Department of Physics, Tsinghua University, Beijing 100084, China, \email{cmsou@mail.tsinghua.edu.cn}
}
%
%
\maketitle

\abstract{Since the original derivation of Hawking radiation, there have been lots of alternative approaches to show the same fact that black holes emit particles as hot bodies with a temperature. These alternative methods generally rely on different conditions and physical quantities to manifest the radiation, providing various points of view of this effect in the intersection of gravity and quantum theory.
This chapter presents some alternative derivations of Hawking radiation in the literature, including the tunneling, anomaly and Green's function methods. From these methods, various features of the black hole system can be seen, such as the gravitational and trace anomalies of the $(1+1)$-dimensional effective theory and the analytical continuation of the complexified spacetime.}


\section{Introduction}
\label{sec:intro}
The original derivation by Hawking \cite{Hawking:1974rv,Hawking:1975vcx} compares the wave modes of quantum field before and after a gravitational collapse of forming a black hole from a star, and the existence of Hawking radiation is reflected by the expectation value of the particle number in each mode, which agrees with thermal radiation with a specific temperature. It was soon after understood that such a state with non-zero particle number exactly corresponds to a thermal state \cite{Parker:1975jm,Wald:1975kc}, and the gravitational collapse can also be replaced by the usual Schwarzschild spacetime to obtain the same result \cite{Unruh:1976db}.\footnote{On the other hand, if we stay with the gravitational collapse, some minimal conditions to derive the radiation have been studied in \cite{Barcelo:2010pj}.} Thus, the conditions required to derive the Hawking radiation are certainly not unique as the original one. Remarkably, the necessary step of this type of derivations of Hawking radiation is the comparison of different vacua, quantified by the Bogoliubov transformation, which has already become a standard technique for studying particle production in curved spacetime \cite{Birrell:1982ix,Ford:1997hb,Parker:2009uva}, with various applications such as cosmology \cite{Ford:2021syk}. Not only can we reconsider the conditions to derive the radiation, but it is also reasonable to consider alternative methods.

Without using the standard technique of quantum field theory in curved spacetime with the Bogoliubov transformation, there are still alternative derivations of Hawking radiation utilizing various features of black holes, such as the tunneling of particles across the horizon determined by energy conservation \cite{Parikh:2004ih}, the anomalies of the effective theory near the horizon highlighting the role of energy-momentum tensor \cite{Christensen:1977jc,Robinson:2005pd} and the emission of particles from the thermal feature of their propagator \cite{Gibbons:1976es}. For the significance of deriving the same result with different methods, it may be a way to see how various properties and points of view of black holes can consistently exist and imply the same effect, because some of them may be related indirectly. In the literature, there is a review including a brief introduction of many alternative derivations \cite{Carlip:2014pma}. Here we choose three types of alternative methods to present and discuss: tunneling, anomaly and Green's function, which are extensively studied in the literature.

This chapter is organized as follows. In section \ref{sec:tunneling}, the tunneling method is introduced, including the Parikh-Wilczek null geodesic method and the Hamilton-Jacobi method. In section \ref{sec:anomaly}, the anomaly method is introduced, starting by deriving the $(1+1)$-dimensional effective theory near the horizon, and the Hawking radiation is derived with the gravitational and trace anomalies in this theory based on some corresponding conditions. In section \ref{sec:Green_fun_method}, we first introduce analytical properties for general thermal Green's functions, and the Hartle-Hawking path integral derivation based on the analyticity of Green function will be demonstrated. Section \ref{sec:conclusion} is the discussion and conclusion.

Here are some conventions used in this chapter. We set the unit to be $G=c=k_B=\hslash=1$, and the metric signature is $(-+++)$. Unless stated explicitly, we denote $\epsilon\to 0^+$ as an infinitesimally small positive number, which will be used in many places in the derivations.

\section{The tunneling method}
\label{sec:tunneling}
Compared to the original Hawking's derivation using quantum field theory in curved spacetime \cite{Hawking:1974rv,Hawking:1975vcx}, the tunneling method emphasizes an intuitive picture of emitting the Hawking radiation: a virtual particle pair created near (inside or outside) the black-hole horizon can materialize if one of them tunnels to another side of the horizon, in which the positive-energy particle in the pair radiates away, whereas the negative-energy particle goes inward and lead to the decrease of black hole's mass. To quantify such a tunneling process, we only need the classical action of the particle in order to calculate the tunneling probability, and this avoids the ambiguity of defining particle states in curved spacetime, showing the advantage and simplicity of this method.

The well-known derivation by Parikh and Wilczek is shown in \cite{Parikh:1999mf}, known as the null geodesic method in the tunneling literature \cite{Vanzo:2011wq}, and some additional discussions by Parikh are in \cite{Parikh:2004rh,Parikh:2004ih}. Besides the derivation by Parikh and Wilczek, the idea of understanding the Hawking radiation as a tunneling process is also seen in \cite{Damour:1976jd} by Damour and Ruffini, and \cite{Srinivasan:1998ty} by Srinivasan and Padmanabhan, in which the latter has been further developed into the Hamilton-Jacobi method of the tunneling \cite{Angheben:2005rm,Padmanabhan:2004tz,Shankaranarayanan:2000gb,Shankaranarayanan:2000qv}. For a more detailed introduction of the tunneling method, there is also a review with various applications to stationary and dynamical black holes, as well as cosmological horizons \cite{Vanzo:2011wq}.

\subsection{The null geodesic method}
We start with the Schwarzschild black hole with the metric
\begin{eqnarray}
    ds^2=-\left(1-\frac{2M}{r}\right)dt^2+\left(1-\frac{2M}{r}\right)^{-1}dr^2+r^2d\Omega^2 \, , \label{eq:Schwarzschild}
\end{eqnarray}
which is well-known to have a coordinate singularity at the horizon $r=2M$, problematic for evaluating the across-horizon tunneling. To avoid the coordinate singularity, the Painlev\'e-Gullstrand coordinates \cite{Painleve:1921,Gullstrand:1922tfa} are introduced by considering a radially free-falling observer with vanishing velocity at infinity, whose geodesic follows the normalized four vector \cite{Kanai:2010ae}
\begin{eqnarray}
    u^\mu=\left(\frac{1}{1-\frac{2M}{r}},-\sqrt{\frac{2M}{r}},0,0\right)\, ,
\end{eqnarray}
which is fixed by the normalization
\begin{eqnarray}
    g_{\mu\nu}u^\mu u^\nu=-1 \, ,
\end{eqnarray}
and the constant of motion (the ratio of energy to rest mass)
\begin{eqnarray}
    -g_{\mu\nu}\xi^\mu u^\nu=1 \, ,
\end{eqnarray}
where $\xi^\mu=(\partial_t)^\mu$ is the timelike Killing vector. We then define a new time $t_p$ such that its constant-time surface is orthogonal to the observer's geodesic, i.e. $\partial_\mu t_p=-u_\mu$, and its total differentiation is
\begin{eqnarray}
        dt_p  &=&  \frac{\partial t_p}{\partial t}dt+\frac{\partial t_p}{\partial r}dr  \\
         &=&  dt+\left(1-\frac{2M}{r}\right)^{-1}\sqrt{\frac{2M}{r}}dr \, ,
\end{eqnarray}
which gives
\begin{eqnarray}
    t_p=t+2\sqrt{2Mr}+2M\log\left(\frac{\sqrt{r}-\sqrt{2M}}{\sqrt{r}+\sqrt{2M}}\right) \, .
\end{eqnarray}
With the new coordinates, the line element becomes regular across the horizon
\begin{eqnarray}
    ds^2=-\left(1-\frac{2M}{r}\right)dt_p^2+2\sqrt{\frac{2M}{r}}dt_p dr+dr^2+r^2d\Omega^2 \, , \label{eq:PG_coordinates}
\end{eqnarray} 
which also has features of time independence and flat Euclidean space on constant-time surfaces.

For the null geodesic method, the energy conservation of the full system of particles and black hole has to be considered because the tunneling ``path" evaluated in this method is defined in the back-reaction sense. As the black hole's mass fluctuates when the particles are emitted, for a spherically symmetric thin shell (s-wave) of particles,\footnote{Besides making the calculation simpler, the case with spherical symmetry circumvent the quantization of gravitons when back reaction is considered \cite{Parikh:2004rh}.} it moves along the geometry (\ref{eq:PG_coordinates}) with $M\to M-\omega$ 
\begin{eqnarray}
    ds^2=-\left[1-\frac{2(M-\omega)}{r}\right]dt_p^2+2\sqrt{\frac{2(M-\omega)}{r}}dt_p dr+dr^2+r^2d\Omega^2 \, , \label{eq:BH_mass_PG}
\end{eqnarray}
which agrees with the Birkhoff's theorem. At the first glance, such a decrease of black hole's mass seems to be a trivial consequence of the energy conservation, but it is necessary to determine the barrier of quantum tunneling \cite{Parikh:2004rh,Parikh:2004ih}. For the usual tunneling process encountered in quantum mechanics, the length of the barrier is determined by classical turning points where the kinetic energy vanishes. If the same point of view is applied to the case of black hole, we would expect that the length of the barrier is infinitesimally small because classical outgoing null geodesics exist just outside the horizon.\footnote{The geodesic equation of outgoing radiation in the geometry (\ref{eq:PG_coordinates}) is $\frac{dr}{dt_p}=1-\sqrt{\frac{2M}{r}}$, which is positive just outside the horizon.} However, with the knowledge that the black hole's mass fluctuates as $M\to M-\omega$, the barrier of quantum tunneling can be understood as the forbidden region from the initial horizon $r=2M$ to the final horizon $r=2(M-\omega)$. In other words, there is no preexisting barrier, but it depends on the particle itself \cite{Parikh:2004ih}.

As a particle near the horizon is infinitely blue-shifted, the wavelength of its wave packet is small enough to consider the geometric optics limit, implying the WKB approximation for the field solution \cite{Parikh:2004rh}
\begin{eqnarray}
    \phi(t_p,r)\approx e^{iS(t_p,r)} \, , \label{eq:WKB_field}
\end{eqnarray}
where $S(t_p,r)$ is the classical action of particle determined by the Hamilton-Jacobi equation. Classically, the particle follows the trajectory determined by initial conditions, and the action is real. Quantum mechanically, it is possible for the particle to pass through some classically forbidden regions, known as the tunneling, and the tunneling probability is evaluated by the imaginary part of the action
\begin{eqnarray}
    \Gamma = e^{-2{\rm Im}\,S} \, . \label{eq:tunneling_prob}
\end{eqnarray}
For a particle pair created just inside the horizon, the positive-energy particle travels from the initial position $r_i=2M$ to the final position $r_f=2(M-\omega)$, and the action is
\begin{eqnarray}
    S=\int^{2(M-\omega)}_{2M} p_r \, dr \, , \label{eq:action_S}
\end{eqnarray}
where $p_r$ is the conjugate momentum of $r$. Without using the explicit solution of $p_r$, the integral can be evaluated with the Hamilton's equations
\begin{eqnarray}
    \frac{dr}{dt_p} &=& \dot{r} \nonumber \\
    &=&\frac{\partial H}{\partial p_r} \nonumber \\
    &=& \frac{d H}{d p_r}\Big|_r  \, ,
\end{eqnarray}
and the imaginary part of (\ref{eq:action_S}) becomes
\begin{eqnarray}
    {\rm Im} S&=& {\rm Im}\int^{2(M-\omega)}_{2M} \int^{p_r}_0 dp'_r \, dr \nonumber \\
    &=& {\rm Im} \int^{2(M-\omega)}_{2M} \int^{M-\omega}_M \frac{dH}{\dot{r}}dr \, .
\end{eqnarray}
With the geodesic equation
\begin{eqnarray}
    \dot{r}&=& 1-\sqrt{\frac{2(M-\omega)}{r}} \, ,
\end{eqnarray}
we can explicitly express ${\rm Im}S$ as a double integral
\begin{eqnarray}
    {\rm Im}S={\rm Im} \int^{2(M-\omega)}_{2M} \int^{M-\omega}_M \frac{dH}{1-\sqrt{\frac{2H}{r}} }dr \, . \label{eq:ImS_double}
\end{eqnarray}
There is a pole when $r=2H$, and the correct way to obtain the imaginary part is to deform the integration contour by the $i\epsilon$ prescription: $H\to H+i\epsilon$ or equivalently $r\to r- i\epsilon$. With the Sokhotski-Plemelj theorem, the integrand becomes
\begin{eqnarray}
    \frac{1}{1-\sqrt{\frac{2H}{r}}-i\epsilon}=i\pi\delta\left(1-\sqrt{\frac{2H}{r}}\right)+{\rm p.v.}\left(\frac{1}{1-\sqrt{\frac{2H}{r}}}\right) \, , \label{eq:iepsilon}
\end{eqnarray}
where ${\rm p.v.}$ is the Cauchy principal value.\footnote{For an integrand $f(x)$ has a singularity at $b$, the Cauchy principal value is ${\rm p.v.} \int_a^c f(x)dx=\lim_{\epsilon->0^+}\left(\int_a^{b-\epsilon}f(x)dx+\int_{b+\epsilon}^c f(x)dx\right)$.} Only the delta function $\delta\left(1-\sqrt{\frac{2H}{r}}\right)$ contributes to the imaginary part, and (\ref{eq:ImS_double}) can be easily evaluated:
\begin{eqnarray}
    {\rm Im}S &=& \pi \int^{2(M-\omega)}_{2M} \int^{M-\omega}_M \delta\left(1-\sqrt{\frac{2H}{r}}\right) dH dr \nonumber \\
    &=& -\pi \int^{2(M-\omega)}_{2M} r \, dr \nonumber \\
    &=& 4\pi\left(M-\frac{\omega}{2}\right)\omega \, . \label{eq:ImS_positive}
\end{eqnarray}
Similarly, the calculation can be done for a particle pair created just outside the horizon in which the negative-energy particle tunnels into the black hole, and we change the null geodesic equation to the correct direction and with the self-gravitational correction \cite{Kraus:1994by}
\begin{eqnarray}
    \dot{r}=-1+\sqrt{\frac{2(M+\omega')}{r}} \, ,
\end{eqnarray}
where $\omega'$ ranges from $0$ to $-\omega$. With the $i\epsilon$ prescription with $\epsilon \to -\epsilon$, the imaginary part of the action is again the same:
\begin{eqnarray}
    {\rm Im} S&=& -{\rm Im} \int^{2(M-\omega)}_{2M} \int^{M-\omega}_M \frac{dH}{1-\sqrt{\frac{2H}{r}} +i\epsilon }dr \nonumber \\
    &=& 4\pi\left(M-\frac{\omega}{2}\right)\omega \, . \label{eq:ImS_negative}
\end{eqnarray}

With the imaginary part of action (\ref{eq:ImS_positive}) and (\ref{eq:ImS_negative}), the tunneling probability (\ref{eq:tunneling_prob}) is 
\begin{eqnarray}
    \Gamma &=& e^{-8\pi\left(M-\frac{\omega}{2}\right)\omega} \nonumber \\
    &=& e^{\Delta S_{\rm B-H}} \, , \label{eq:Gamma_S_B_H}
\end{eqnarray}
which is related to the change of the Bekenstein-Hawking entropy. For the leading order with $\omega \ll M$, (\ref{eq:Gamma_S_B_H}) reduces to the Boltzmann factor, and the Hawking temperature is recovered
\begin{eqnarray}
    T_{ H}=\frac{1}{8\pi M}\, .
\end{eqnarray}
For the next order $O(\omega^2)$ in (\ref{eq:Gamma_S_B_H}), it originates from the energy conservation that emitting radiation reduces the black hole's mass, and hence increasing its temperature.

As a final remark of the null geodesic method, there are two conditions required during the derivation which should have a better justification:
\begin{svgraybox}
    \begin{itemize}
        \item Back-reaction by energy conservation in (\ref{eq:BH_mass_PG}) is necessary to define the tunneling path, but there are other classically forbidden paths defined in a way not relying on the back-reaction.
        \item The sign of $i\epsilon$ prescription in (\ref{eq:iepsilon}) should be justified with the analytical properties of $S$ since flipping the sign $\epsilon\to-\epsilon$ leads to minus the imaginary part.
    \end{itemize}
\end{svgraybox} 
These two aspects will be addressed with the Hamiltonian-Jacobi method and analytical continuation in section \ref{sec:HJ_method}.

\subsection{The Hamiltonian-Jacobi method}\label{sec:HJ_method}
This method directly applies the WKB approximation (\ref{eq:WKB_field}) to the massless Klein-Gordon equation
\begin{eqnarray}
    -\frac{1}{\sqrt{-g}}\partial_\mu\left(\sqrt{-g}g^{\mu\nu}\partial_\nu\phi\right)=0 \, ,
\end{eqnarray}
and thus the action $S(x^\mu$) satisfies the relativistic Hamiltonian-Jacobi equation \cite{Angheben:2005rm} 
\begin{eqnarray}
    g^{\mu\nu}\partial_\mu S \partial_\nu S=0 \, . \label{eq:H-J_eq}
\end{eqnarray}
This method has been widely applied to generalize the tunneling process to complex paths by analytical continuation of coordinates \cite{Padmanabhan:2004tz,Shankaranarayanan:2000gb,Shankaranarayanan:2000qv,Srinivasan:1998ty}. Firstly, we follow \cite{Vanzo:2011wq} to repeat the calculation for a null geodesic from just inside the horizon to just outside, but without considering back-reaction, and we will see that the result agrees with the previous calculation. The Hamilton-Jacobi equation (\ref{eq:H-J_eq}) in the Painlev\'e-Gullstrand coordinates (\ref{eq:PG_coordinates}) is
\begin{eqnarray}
    &-&\left(\partial_{t_p} S\right)^2+2\sqrt{\frac{2M}{r}}\partial_{t_p}S \partial_r S+\left(1-\frac{2M}{r}\right)\left(\partial_r S\right)^2 \nonumber \\
    &=&-\omega^2-2\omega \sqrt{\frac{2M}{r}}\partial_r S +\left(1-\frac{2M}{r}\right)\left(\partial_r S\right)^2 \nonumber \\ \label{eq:HJE_PG}
    &=& 0 \, ,
\end{eqnarray}
where the constant energy $\omega=-\partial_{t_p}S$, and the angular part is neglected, so we solve for $\partial_r S$ with the correct sign for outgoing
\begin{eqnarray}
    \partial_r S=\frac{\omega}{1-\sqrt{\frac{2M}{r}}} \, . \label{eq:d_rS}
\end{eqnarray}
For a null geodesic near the horizon
\begin{eqnarray}
    ds^2&=&2dt_p dr+dr^2 \nonumber \\
    &=&0 \, ,
\end{eqnarray}
we have $dt_p=-\frac{dr}{2}$, and thus the imaginary part of the action with (\ref{eq:d_rS}) is
\begin{eqnarray}
    {\rm Im}S &=& {\rm Im}\int \left(\partial_r S dr+\partial_{t_p}S dt_p\right) \nonumber \\
    &=& {\rm Im}\int^{2M+\epsilon}_{2M-\epsilon} \frac{\omega}{1-\sqrt{\frac{2M}{r}}-i\epsilon}  dr  \nonumber \\
    &=& 4\pi M \omega \, , \label{eq:integral_HJ}
\end{eqnarray}
where the real temporal part does not contribute any imaginary value, and the $i\epsilon$ prescription $r\to r-i\epsilon$ is applied. Compared to (\ref{eq:ImS_negative}), this agrees with the null geodesic method when the back-reaction term is neglected, suggesting that we can also have a tunneling process without considering any back-reaction, in which the path is the classically-forbidden outgoing null geodesic. In fact, the path in the integral (\ref{eq:integral_HJ}) can be extended to $r_1<r<r_2$ where $r_1<2M$ and $r_2>2M$ are inside and outside the horizon respectively because the contributions in $[r_1,2M-\epsilon]$ and $[2M+\epsilon,r_2]$ are real, so general null geodesics from inside to outside the horizons give the same tunneling probability.

\begin{figure}
    \centering
    \includegraphics[width=0.7\textwidth]{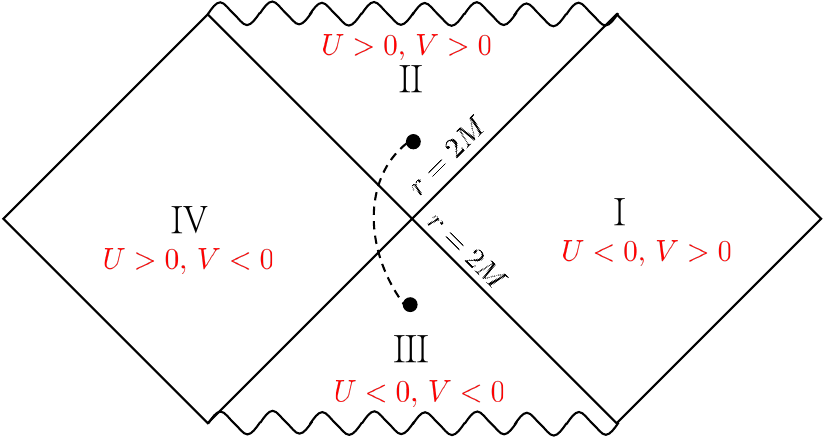}
    \caption{A Schwarzschild black hole with the Kruskal extension. The dashed path indicates the complex path (\ref{eq:complex_path}) from the region II to III. \label{fig:complex_path}}
\end{figure}

On the other hand, the sign of $i \epsilon$ prescription can be justified if the tunneling path is analytically continued to a complex path. With the Kruskal extension, a Schwarzschild black hole can be separated into four regions, including outside the horizon (region I), inside the horizon (region II) and inside the horizon and with time reversal (region III), as shown in figure \ref{fig:complex_path}. As demonstrated in \cite{Vanzo:2011nd}, instead of directly tunneling from region II to I, there exists a complex path tunneling from region II to III and then following the classically allowed path from region III to I, described by the complex rotation of the Kruskal coordinates with $t\to t-4Mi\theta$:
\begin{eqnarray}
    \tilde{U}&=&e^{i\theta}\left(1-\frac{r}{2M}\right)^{\frac{1}{2}}e^{\frac{r-t}{4M}} \nonumber \\
    &=& e^{i\theta} U \nonumber \\
    \tilde{V}&=&e^{-i\theta}\left(1-\frac{r}{2M}\right)^{\frac{1}{2}}e^{\frac{r+t}{4M}}\nonumber \\
    &=& e^{-i\theta} V \, , \label{eq:complex_path}
\end{eqnarray}
where $\theta=0$ and $\theta=\pi$ correspond to the region II and III respectively. 
With the Kruskal coordinates, the action integral $\int\partial_\mu S dx^\mu$ is expressed from $\int \left(\partial_r S dr + \partial_{t_p} S d{t_p}\right)$ to $\int \left(\partial_{\tilde{U}}S d\tilde{U}+\partial_{\tilde{V}}S d\tilde{V}\right)$, and the imaginary part of the action is 
\begin{eqnarray}
    {\rm Im}S &=& {\rm Im} \int_{II\to III} \left(\partial_{\tilde{U}}S d\tilde{U}+\partial_{\tilde{V}}S d\tilde{V}\right) \nonumber \\
    &=&{\rm Im}\left[i\int_0^\pi d\theta \, \left(U\partial_{U}S -V\partial_{V}S \right)\right] \nonumber\\
    &=& -4M\int_0^\pi d\theta \, \partial_t S|_r \nonumber \\
    &=& -4M\int_0^\pi d\theta \, \partial_{t_p} S \nonumber \\
    &=& 4\pi M \omega \, . \label{eq:ImS_UV}
\end{eqnarray}
It is also easy to check that the following path from region III to I, described by the time-reversed (equivalently ingoing) one from (\ref{eq:HJE_PG})
\begin{eqnarray}
    \partial_r S=-\frac{\omega}{1+\sqrt{\frac{2M}{r}}} \, ,
\end{eqnarray}
is classically allowed since there is no any singularity at $r=2M$. Therefore, the imaginary part contributed by the complex path (\ref{eq:ImS_UV}), obtained by the imaginary change of the Schwarzschild time $t$, agrees with (\ref{eq:ImS_negative}) (ignore the back-reaction term) and (\ref{eq:integral_HJ}), suggesting that the $i\epsilon$ prescription makes sense.\footnote{Certainly, a better justification should connect the complex continuations between $t$ and $r$, and the demonstration here is useful to determine the sign of the $i\epsilon$ prescription.} Such an analytical continuation to the region III was also applied in the Hartle-Hawking path-integral derivation of Hawking radiation \cite{Hartle:1976tp}, in which the analytical properties of Green function is studied, suggesting that the choice of complex path (\ref{eq:complex_path}) is reasonable. We will discuss this derivation in detailed in section \ref{sec:path_integral_approach}, and the same tunneling probability is obtained.

\section{The anomaly method}
\label{sec:anomaly}
The relationship between anomaly and Hawking radiation has been noticed by Christensen and Fulling in 1977, soon after Hawking's discovery \cite{Christensen:1977jc}, but the result relies on the trace anomaly of the energy-momentum tensor for a field with conformal invariance in $1+1$ dimension, limiting the applicability of this method in real cases and general spacetimes. A more feasible anomaly method was shown much later by Robinson and Wilczek in 2005 \cite{Robinson:2005pd}, in which the anomaly in general covariance, the gravitational anomaly, was applied to derive the Hawking radiation. Soon after the original paper by Robinson and Wilczek, the gravitational-anomaly method was generalized to Reissner-Nordstr\"om \cite{Iso:2006wa} and Kerr \cite{Iso:2006ut,Murata:2006pt} black holes.

The detailed derivation of black hole effective field theory is demonstrated in Robinson's thesis \cite{Robinson:2005ph}. The recipe is as follows:\footnote{For the original paper by Robinson and Wilczek \cite{Robinson:2005pd}, the outgoing modes near the horizon are discarded, and see the remark at the end of section \ref{sec:gravitaitonal_anomaly} for a detailed discussion.}
\begin{svgraybox}
    Near the horizon, i.e. $0\leq r-r_h \leq \epsilon$
    \begin{itemize}
        \item A scalar field can be described by $(1+1)$-dimensional theory in the $r-t$ section with the partial wave decomposition.
        \item Ingoing modes near the horizon are discarded because they cannot affect the effective theory outside the horizon, leading to a $(1+1)$-dimensional chiral scalar field.
        \item Such a chiral scalar field has the gravitational anomaly in its energy-momentum tensor, and the full quantum theory must admit a flux to compensate this. This flux will be the Hawking radiation.
    \end{itemize}
\end{svgraybox}

\subsection{The effective $(1+1)$-dimensional massless theory near the horizon}
We consider a scalar field which generally has mass and interaction terms in its action
\begin{eqnarray}
    S &=& -\frac{1}{2}\int d^4x \sqrt{-g}\left(g^{\mu\nu}\partial_\mu\phi\partial_\nu\phi+\sum_{n=2}^\infty\lambda_n\phi^n\right) \nonumber \\
    &=& \frac{1}{2}\int d^4x \sqrt{-g}\left[\frac{1}{\sqrt{-g}}\phi \partial_\mu\left(\sqrt{-g}g^{\mu\nu}\partial_\nu\phi\right)-\sum_{n=2}^\infty\lambda_n\phi^n\right] \nonumber \\
    &=&\frac{1}{2}\int d^4x \sqrt{-g}\left(\phi g^{\mu\nu}\nabla_\mu \nabla_\nu \phi-\sum_{n=2}^\infty \lambda_n\phi^2\right)
    \, , \label{eq:scalar_action}
\end{eqnarray}
where we did an integration by parts in the second line, $\nabla_\mu$ is the covariant derivative, and $\lambda_2=m^2$ is the mass term. For the simplest case, the metric is static and spherically symmetric
\begin{eqnarray}
    ds^2=-f(r)dt^2+\frac{1}{f(r)}dr^2+r^2d\Omega^2 \, , \label{eq:metric_f}
\end{eqnarray}
so the Schwarzschild black hole (\ref{eq:Schwarzschild}) corresponds to $f(r)=1-\frac{2M}{r}$, and the horizon is thus determined by $f(r_H)=0$ as well as the surface gravity 
\begin{eqnarray}
    \kappa=\frac{1}{2}f'(r_H) \, ,
\end{eqnarray}
is finite. For making the following calculation simple, the tortoise coordinates are introduced
\begin{eqnarray}
    \frac{dr}{dr_*}&=&f(r) \, , \\
    ds^2 &=& f(r)(-dt^2+dr_*^2)+r^2d\Omega^2 \, ,
\end{eqnarray} 
and the variable $r_*$ near the horizon is
\begin{eqnarray}
    r_*&\approx& \frac{1}{f'(r_H)}\log\left(\frac{r-r_H}{r_0-r_H}\right) \nonumber \\
    &=& \frac{1}{2\kappa}\log\left(\frac{r-r_H}{r_0-r_H}\right)  \, , \label{eq:rstar_approx}
\end{eqnarray}
where $r_0$ is an unimportant integration constant, and therefore $r_*\to-\infty$ when $r\to r_H$.
The action (\ref{eq:scalar_action}) in this coordinate system is
\begin{eqnarray}
    S&=&\frac{1}{2}\int dt dr_* d^2\Omega \,\Bigg[\phi\left(-r^2\partial_t^2+r^2\partial_{r_*}^2+2rf\partial_{r_*}-f L^2-fm^2r^2\right)\phi \nonumber \\
    &-&\sum_{n=3}^\infty \lambda_n f r^2 \phi^n \Bigg] \, , \label{eq:action_in_metric}
\end{eqnarray}
where 
\begin{eqnarray}
    L^2=-\frac{1}{\sin\theta}\partial_\theta\left(\sin\theta\partial_\theta\right)-\frac{1}{\sin^2\theta}\partial^2_\varphi \, , \label{eq:Lsquare}
\end{eqnarray}
is the orbital angular momentum.

In the static and spherically symmetric metric, the field has a mode expansion
\begin{align}
    \phi=\frac{1}{r}\sum_{l,m}\varphi_{lm}(t,r_*)Y_{lm}(\Omega) \, ,
\end{align}
where $Y_{lm}(\Omega)$ is the spherical harmonics, and the action (\ref{eq:action_in_metric}) can be further simplified by integrating over the angles
\begin{eqnarray}
    S&=&\frac{1}{2}\int dtdr_* \, \Bigg[\sum_{l,m'}\varphi_{lm'}\left(-\partial_t^2+\partial^2_{r_*}-\frac{ff'}{r}-f\frac{l(l+1)}{r^2}-fm^2\right)\varphi_{lm'} \nonumber \\
    &-&\sum_{n=3}^\infty \frac{\lambda_n f}{r^{n-2}}\sum_{l_1,\dots,l_n,m_1,\dots,m_n} {^n}C^{\{{\bf m}\}}_{\{{\bf l} \}}\varphi_{l_1m_1}\dots \varphi_{l_nm_n} \Bigg]\, , \label{eq:action_modes_expand}
\end{eqnarray}
where the $l(l+1)$ term is obtained from (\ref{eq:Lsquare})
\begin{eqnarray}
    L^2Y_{lm}=l(l+1)Y_{lm} \, ,
\end{eqnarray}
and
\begin{eqnarray}
    \int d^2\Omega \, Y_{lm}(\Omega) Y_{l'm'}(\Omega)&=& \delta_{l,l'}\delta_{m,m'} \\
    {^n}C^{\{{\bf m}\}}_{\{{\bf l} \}}&=&\int d^2\Omega \, Y_{l_1m_1}(\Omega)\dots Y_{l_nm_n} \, ,
\end{eqnarray}
are the integrals of the product of $n$ spherical harmonics. With (\ref{eq:rstar_approx}), the function $f(r)$ near the horizon is exponentially suppressed in $r_*\to-\infty$
\begin{eqnarray}
    f(r_*)&\approx& f'(r_H)(r-r_H) \nonumber \\
    &\approx& 2\kappa(r_0-r_H)e^{2\kappa r_*} \, ,
\end{eqnarray} 
and thus the terms with $f(r)$ in the action (\ref{eq:action_modes_expand}) can be neglected, leading to an effectively free $(1+1)$-dimensional massless field near the horizon in the $r-t$ section
\begin{eqnarray}
    S\approx \frac{1}{2}\int dtdr_* \, \sum_{lm}\varphi_{lm}\left(-\partial_t^2+\partial_{r_*}^2\right)\varphi_{lm} \, , \label{eq:1p1_effective}
\end{eqnarray}
where we can simply neglect the subscript $lm$.

The equation of motion from (\ref{eq:1p1_effective}) is just the usual wave equation
\begin{eqnarray}
    \left(\partial_t^2-\partial_{r_*}^2\right)\varphi&=& \partial_u\partial_v \varphi \nonumber \\
    &=& 0 \, ,
\end{eqnarray}
where the variables
\begin{eqnarray}
    u&=&t-r_*  \nonumber \\
    v&=&t+r_* \, .
\end{eqnarray}
It is clear that the field near the horizon can have two independent sets of solutions: the outgoing and ingoing modes with the forms 
\begin{eqnarray}
    \varphi(u)&=&\frac{e^{-i\omega u}}{\sqrt{4\pi\omega}} \nonumber \\
    \varphi(v)&=&\frac{e^{-i\omega v}}{\sqrt{4\pi\omega}} \, , \label{eq:mode_fun}
\end{eqnarray}
respectively, which play roles like right- and left-handed fields.

\subsection{A simple derivation of the trace and (covariant) gravitational anomalies of the effective theory}
    We follow \cite{Brout:1995rd} to give a simple derivation of the trace anomaly of the effective theory (\ref{eq:trace_anomaly}), and a more detailed discussion can be found in \cite{Davies:1976ei,Davies:1977}. For the $(1+1)$-dimensional metric (\ref{eq:metric_rt}), we can rewrite it with null coordinates
    \begin{eqnarray}
        ds^2=-C(U,V)dUdV \, , \label{eq:CdUdV}
    \end{eqnarray}
    where for example $U=u=t-r_*$ and $V=v=t+r_*$ corresponds to $C=f$, and we can further rescale $U$ and $V$ to get the form in (\ref{eq:CdUdV}). The components of the energy-momentum tensor of the field is
    \begin{eqnarray}
        T_{UU}=\partial_U \varphi\partial_U \varphi \, , \, T_{VV}=\partial_V \varphi\partial_V \varphi \, ,
    \end{eqnarray}
    and classically we will get $T_{UV}=0$ which agrees with the vanishing trace.

    As a first attempt, the quantum expectation value of $T_{UU}$ can be evaluated as follows
    \begin{eqnarray}
        \langle T_{UU} \rangle &=& \lim_{U\to U'} \partial_U \partial_{U'} \langle \varphi (U) \varphi (U') \rangle \nonumber \\
        &=& -\frac{1}{4\pi} \lim_{U\to U'} \partial_U \partial_{U'} \log\left|U-U'\right| \nonumber \\
        &=& -\frac{1}{4\pi} \lim_{U\to U'} \frac{1}{(U-U')^2} \, , \label{eq:divergent_TUU}
    \end{eqnarray} 
    where the two-point function is calculated with the mode functions (\ref{eq:mode_fun})
    \begin{eqnarray}
        \langle \varphi (U) \varphi (U') \rangle &=& \frac{1}{4\pi}\lim_{L\to \infty}\int^{+\infty}_{L^{-1}}\frac{d\omega}{\omega} e^{-i\omega(U-U')} \nonumber \\
        &=& -\frac{\log\left|U-U'\right|}{4\pi}+\lim_{L\to \infty}\frac{1}{4\pi}\left(-\gamma_E-\frac{i\pi}{2}+\log L\right) \, ,
    \end{eqnarray}
    in which the second constant term cannot contribute after taking derivatives. The evaluation of $\langle T_{VV} \rangle$ can be done similarly by replacing $U$ to  $V$. To resolve the divergence in (\ref{eq:divergent_TUU}) when $U\to U'$, we have to renormalize it by subtracting the one by local inertial modes, in which the inertial coordinates $(\hat{U},\hat{V})$ are defined by
    \begin{eqnarray}
        \hat{U}-\hat{U}_0 &=& \int^U_{U_0} \frac{C(U',V_0)}{C(U_0,V_0)}dU' \nonumber \\
        \hat{V}-\hat{V}_0 &=& \int^V_{V_0} \frac{C(U_0,V')}{C(U_0,V_0)}dV' \, ,
    \end{eqnarray} 
    which can be interpreted as the affine parameters of light-like geodesics. In these coordinates, the inertial modes are
    \begin{eqnarray}
        \tilde{\varphi}(\hat{U})&=&\frac{e^{-i\omega \hat{U}}}{\sqrt{4\pi\omega}} \nonumber \\
        \tilde{\varphi}(\hat{V})&=&\frac{e^{-i\omega \hat{V}}}{\sqrt{4\pi\omega}} \, ,
    \end{eqnarray}
    so the renormalized energy-momentum tensor is
    \begin{eqnarray}
        \langle T_{UU}\rangle_{\rm ren}&=&-\frac{1}{4\pi} \lim_{U\to U'}\partial_U\partial_{U'}\left(\log|U-U'|-\log|\hat{U}(U)-\hat{U}(U')|\right) \nonumber \\
        &=& -\frac{1}{12\pi}\sqrt{\partial_U\hat{U}}\partial^2_U \left(\frac{1}{\sqrt{\partial_U \hat{U}}}\right) \nonumber \\
        &=& -\frac{1}{12\pi} C^{\frac{1}{2}} \partial^2_U C^{-\frac{1}{2}} \, , \label{eq:T_UU}
    \end{eqnarray}
    and similarly we have 
    \begin{eqnarray}
        \langle T_{VV}\rangle_{\rm ren}&=& -\frac{1}{12\pi} C^{\frac{1}{2}} \partial^2_V C^{-\frac{1}{2}} \, . \label{eq:T_VV}
    \end{eqnarray}
    The $T_{UV}$ component is obtained by requiring the conservation
    \begin{eqnarray}
        \nabla_\nu \langle T^{\nu}_U \rangle_{\rm ren}&=& C^{-1}\partial_V \langle T_{UU}\rangle_{\rm ren}+ \partial_U \left(C^{-1}\langle T_{UV}\rangle_{\rm ren}\right) \nonumber \\
        &=& 0 \, ,
    \end{eqnarray}
    which gives the $T_{UV}$ component
    \begin{eqnarray}
        \langle T_{UV}\rangle_{\rm ren}&=&-\frac{1}{24\pi}\partial_U\partial_V\log C \, , \label{eq:T_UV} 
    \end{eqnarray}

    From here onwards, we can neglect $\langle \dots \rangle_{\rm ren}$ for simplicity, and $T_{\mu\nu}$ should be understood as the quantum expectation value. Combining (\ref{eq:T_UU}), (\ref{eq:T_VV}) and (\ref{eq:T_UV}), the energy-momentum tensor can be collectively expressed as \cite{Davies:1976ei}
    \begin{eqnarray}
        T_{\mu\nu}= \frac{R}{48\pi}g_{\mu\nu}+\theta_{\mu\nu}\, , \label{eq:T_munu_theta}
    \end{eqnarray}
    where the Ricci scalar $R$ is
    \begin{eqnarray}
        R&=& 4C^{-1}\partial_U\partial_V\log C \, ,
    \end{eqnarray}
    and the components of  $\theta_{\mu\nu}$ are
    \begin{eqnarray}
        \theta_{UU}&=&-\frac{1}{12\pi} C^{\frac{1}{2}} \partial^2_U C^{-\frac{1}{2}} \nonumber \\
        \theta_{VV}&=&-\frac{1}{12\pi} C^{\frac{1}{2}} \partial^2_V C^{-\frac{1}{2}} \nonumber \\
        \theta_{UV}&=&\theta_{VU}=0 \, . \label{eq:theta_components}
    \end{eqnarray}
    From the conservation of $T_{\mu\nu}$, we also have
    \begin{eqnarray}
        \nabla^\mu \theta_{\mu\nu}=-\frac{1}{48\pi}\partial_\nu R \, . \label{eq:divergence_theta}
    \end{eqnarray}
    \begin{svgraybox}
    From (\ref{eq:T_munu_theta}), it is clear that the energy-momentum tensor has a trace anomaly 
        \begin{eqnarray}
            T^\alpha_{\alpha} &=&  \frac{1}{24\pi} R  \, . \label{eq:trace_anomaly}
        \end{eqnarray}
    \end{svgraybox}

    Next we follow a simple argument in \cite{Fulling:1986rk,Banerjee:2008sn} to show the (covariant) gravitational anomaly for the chiral theory obtained by neglecting ingoing modes. As $U$ and $V$ are null variables representing outgoing (right-handed) and ingoing (left-handed) modes, (\ref{eq:theta_components}) implies that we can separate $\theta_{\mu\nu}$ into two parts
    \begin{eqnarray}
        \theta_{\mu\nu}&=&\theta_{\mu\nu}^{(R)}+\theta_{\mu\nu}^{(L)} \nonumber \\
        \theta_{UU}^{(R)}&=&\theta_{UU} \nonumber \\
        \theta_{VV}^{(L)}&=&\theta_{VV} \nonumber \\
        \theta_{VV}^{(R)}&=&\theta_{UV}^{(R)}=\theta_{UU}^{(L)}=\theta_{UV}^{(L)}=0 \, .
    \end{eqnarray}
    With this separation, the energy-momentum tensor is also separated as $T_{\mu\nu}=T^{(R)}_{\mu\nu}+T^{(L)}_{\mu\nu}$ where
    \begin{eqnarray}
        T^{(R/L)}_{\mu\nu}&=&\frac{R}{96\pi}g_{\mu\nu}+\theta^{(R/L)}_{\mu\nu} \, ,
    \end{eqnarray}
    and for the chiral theory with only the outgoing (right-handed) modes, $T^{(R)}_{\mu\nu}$ is not conserved:
    \begin{eqnarray}
        \nabla^\mu T_{\mu \nu}^{(R)}&=&\frac{\partial_\nu R}{96\pi}+\nabla^\mu\theta^{(R)}_{\mu\nu} \nonumber \\
        &=&\frac{\partial_\nu R}{96\pi}+\delta^U_\nu\nabla^\mu\theta_{\mu U} \nonumber \\
        &=& \begin{cases}
            -\frac{\partial_U R}{96\pi} & \nu=U \\
            \frac{\partial_V R}{96\pi}  & \nu=V 
        \end{cases} \, ,
    \end{eqnarray}
    where (\ref{eq:divergence_theta}) is used, and this can collectively be rewritten as 
    \begin{svgraybox}
        \begin{eqnarray}
            \nabla^\mu T_{\mu \nu}^{(R)} &=& \frac{1}{96\pi} \epsilon_{\mu \nu} \partial^\mu R \, ,  \label{eq:gravitational_anomaly_R}
        \end{eqnarray}
    where $\epsilon_{\mu\nu}$ is the Levi-Civita tensor. This is the covariant gravitational anomaly in this chiral theory.
    \end{svgraybox}

\subsection{Hawking radiation by resolving the gravitational anomaly}\label{sec:gravitaitonal_anomaly}
Since the ingoing modes are not relevant to the effective theory outside the horizon, they should be discarded, leading to a $(1+1)$-dimensional chiral scalar theory. Generally, such a chiral theory is known to have a (consistent) gravitational anomaly in the energy-momentum tensor, obtained from loop diagrams \cite{Alvarez-Gaume:1983ihn,Bertlmann:2000da}
\begin{eqnarray}
    \nabla_\mu T_{(\chi)} {^\mu_\nu}&=&\frac{1}{\sqrt{-g_{(2)}}}\partial_\mu N^\mu_\nu \nonumber \\
    &\equiv& A_\nu\, , \label{eq:anomaly_chiral}
\end{eqnarray}
where $g_{(2)}{_{\mu\nu}}$ is the $(1+1)$-dimensional metric of the $r-t$ section of (\ref{eq:metric_f})
\begin{eqnarray}
    ds^2=-f(r)dt^2+\frac{1}{f(r)}dr^2 \, , \label{eq:metric_rt}
\end{eqnarray}
and
\begin{eqnarray}
    N^\mu_\nu&=&\eta \frac{1}{96\pi}\epsilon^{\beta \mu}\partial_\alpha \Gamma^{\alpha}_{\nu \beta} \, , \label{eq:N_mu_nu}
\end{eqnarray}
in which the convention of the Levi-Civita tensor is $\epsilon^{01}=-\epsilon^{10}=1$. Here we denote $\eta$ as a sign discrepancy appeared in the literature, such as $\eta=1$ used in \cite{Iso:2006ut,Iso:2006wa,Iso:2006xj}, whereas $\eta=-1$ used in \cite{Murata:2006pt}, and we will show that this affects the sign of $T^r_t$ for the flux of Hawking radiation.

On the other hand, there is also a covariant version of the gravitational anomaly, previously derived in (\ref{eq:gravitational_anomaly_R}) from a relatively simple set up\footnote{In \cite{Iso:2006wa}, there is also a $\sqrt{-g_{(2)}}$ in the denominator (\ref{eq:anomaly_covariant}), but this does not affect the result since it is equal to $1$.} 
\begin{eqnarray}
    \nabla_\mu \tilde{T}_{(\chi)}{^\mu_\nu}&=&-\eta\frac{1}{96\pi }\epsilon_{\mu \nu}\partial^\mu R \nonumber \\
    &\equiv& \tilde{A}_\nu \, , \label{eq:anomaly_covariant}
\end{eqnarray}
where $\tilde{T}_{(\chi)}{^\mu_\nu}$ is obtained by adding the Bardeen-Zumino polynomial to $T_{(\chi)}{^\mu_\nu}$ \cite{Bardeen:1984pm}, and $R$ is the Ricci scalar of the metric (\ref{eq:metric_rt}),\footnote{The Bardeen-Zumino polynomial is basically determined by the right-hand sides of (\ref{eq:anomaly_chiral}) and (\ref{eq:anomaly_covariant}), so we do not write it down explicitly.} and $\epsilon_{01}=-\epsilon_{10}=-1$ by lowering indices. In the literature such as \cite{Iso:2006wa}, the consistent version (\ref{eq:anomaly_chiral}) is commonly used to fix the form of the flux of Hawking radiation, and the covariant version (\ref{eq:anomaly_covariant}) is used to fix a boundary condition at the horizon in the final step. We choose to follow this derivation. On the other hand, directly using the covariant version also leads to the correct result \cite{Banerjee:2007qs}.

In the $r-t$ section, the non-vanishing components of the Christoffel symbol $\Gamma^{\alpha}_{\nu \beta}$ for the metric (\ref{eq:metric_f}) are 
\begin{eqnarray}
    \Gamma^{t}_{tr}&=&\Gamma^{t}_{rt}=\frac{f'}{2f} \, ,  \label{eq:Christoffel_t}\\
    \Gamma^{r}_{tt}&=&\frac{ff'}{2} \, , \, \Gamma^{r}_{rr}=-\frac{f'}{2f} \, , \label{eq:Christoffel_r} 
\end{eqnarray}
and thus the explicit components of $N^\mu_\nu$ in the $r-t$ section can thus be calculated with (\ref{eq:N_mu_nu})
\begin{eqnarray}
    N^r_t &=& \eta\frac{1}{192\pi}\left(f'^2+f'' f\right) \nonumber \\
    N^t_r &=& -\eta\frac{1}{192\pi f^2}\left(f'^2-f'' f\right) \nonumber \\
    N^t_t&=&0 \nonumber \\
    N^r_r&=&0 \, , \label{eq:N_components}
\end{eqnarray}
where the trick to simplify the calculation is that all the components of $\Gamma^\alpha_{\nu\beta}$ are $t$-independent, so only the spatial derivatives $\partial_r\Gamma^{r}_{\nu\beta}$ in (\ref{eq:N_mu_nu}) contribute, which involve the two components in (\ref{eq:Christoffel_r}). With $\Gamma^\alpha_{\nu\beta}$, the $A_\nu$ in (\ref{eq:anomaly_chiral}) is obtained easily
\begin{eqnarray}
    A_t&=& \eta\frac{1}{192\pi}\left(f'^2+f'' f\right)' \nonumber \\
    A_r&=& 0 \, . \label{eq:Atr}
\end{eqnarray}
Similarly, with the Ricci scalar $R$, we can determine $\tilde{A}_\mu$ in (\ref{eq:anomaly_covariant})
\begin{eqnarray}
    R&=&-f'' \nonumber \\
    \tilde{A}_t &=& \eta\frac{1}{192\pi}\left(2ff''-f'^2\right)' \nonumber \\
    \tilde{A}_r &=& 0 \, . \label{eq:Atildetr}
\end{eqnarray}
For the static metric, $T_{(\chi)}{^\mu_\nu}$ is time-independent, and (\ref{eq:anomaly_chiral}) reduced to analytically solvable ordinary differential equations
\begin{eqnarray}
    \partial_r T_{(\chi)}{^r_t} &=&A_t(r) \nonumber \\
    \partial_r T_{(\chi)}{^r_r}+\frac{f'}{2f}\left(2T_{(\chi)}{^r_r}-T_{(\chi)}{^\alpha_\alpha}\right) &=& 0 \, ,
\end{eqnarray}
where $T_{(\chi)}{^\alpha_\alpha}=\partial_r T_{(\chi)}{^t_t}+\partial_r T_{(\chi)}{^r_r}$ is the trace of energy-momentum tensor. The solutions of the differential equations fix all the four components of $T_{(\chi)}{^{\mu}_\nu}$
\begin{eqnarray}
    T_{(\chi)}{^r_t}&=& -K_\chi+ \int^r_{r_H}A_t(r')dr' \nonumber \\
    &=&  -f^2 T_{(\chi)}{^t_r} \nonumber \\
    T_{(\chi)}{^r_r}&=&\frac{K_\chi+Q_\chi}{f}+\frac{1}{2f}\int^r_{r_H}T_{(\chi)}{^\alpha_\alpha}(r')f'(r')dr' \nonumber \\
    &=& T_{(\chi)}{^\alpha_\alpha}-T_{(\chi)}{^t_t} \, , \label{eq:T_chi}
\end{eqnarray}
where $K_\chi$ and $Q_\chi$ are integration constants.

Now we take a closer look to the energy-momentum tensor of the chiral theory with neglecting ingoing modes near the horizon. Besides the anomalous part in $r_H\leq r \leq r_H+\epsilon$, the energy-momentum tensor conserves as usual outside the near-horizon region ($r>r_H+\epsilon$), so $T^\mu_\nu$ is separated into two parts 
\begin{eqnarray}
    T^\mu_\nu&=& \begin{cases}
    T_{(\chi)}{^\mu_\nu}\, , & r_H\leq r \leq r_H+\epsilon\\
    T_{(o)}{^\mu_\nu}\, , & r>r_H+\epsilon
    \end{cases} \nonumber \\
    &=&T_{(\chi)}{^\mu_\nu}H+ T_{(o)}{^\mu_\nu} \Theta_+ \, , \label{eq:Tmunu_separated}
\end{eqnarray}
where the Heaviside step functions $\Theta_+=\Theta(r-r_H-\epsilon)$, and $H=1-\Theta_+$. The divergences of the two parts of $T^\mu_\nu$ are 
\begin{eqnarray}
    \nabla_\mu \left(T_{(\chi)}{^\mu_\nu} H\right)&=& \nabla_\mu T_{(\chi)} {^\mu_\nu}H+T_{(\chi)}{^\mu_\nu}\partial_\mu H \nonumber \\
    &=& \frac{1}{\sqrt{-g_{(2)}}} \partial_\mu N^\mu_\nu H +T_{(\chi)}{^\mu_\nu} \partial_\mu H \nonumber \\
    &=& \frac{1}{\sqrt{-g_{(2)}}} \partial_\mu\left(N^\mu_\nu H\right)+\left(T_{(\chi)}{^\mu_\nu}-\frac{1}{\sqrt{-g_{(2)}}}N^{\mu}_\nu\right) \partial_\mu H \nonumber \\
    &=&\partial_\mu\left(N^\mu_\nu H\right)-\left(T_{(\chi)}{^\mu_\nu}-N^{\mu}_\nu\right)\partial_\mu \Theta_+ \, , \label{eq:divergence_near}
\end{eqnarray}
where (\ref{eq:anomaly_chiral}) and $\sqrt{-g_{(2)}}=1$ are applied, and 
\begin{eqnarray}
    \nabla_\mu\left( T_{(o)}{^\mu_\nu}\Theta_+ \right) &=& T_{(o)}{^\mu_\nu} \partial_\mu \Theta_+ 
    + \nabla_\mu T_{(o)}{^\mu_\nu}\Theta_+  \nonumber \\
    &=& T_{(o)}{^\mu_\nu} \partial_\mu \Theta_+  \, , \label{eq:divergence_outside}
\end{eqnarray}
where the conservation outside the near-horizon region is applied
\begin{eqnarray}
    \nabla_\mu T_{(o)}{^\mu_\nu}=0 \, .
\end{eqnarray}
The conservation in $r>r_H+\epsilon$ also fixes the form of $T_{(o)}{^\mu_\nu}$, which is just the homogeneous version of (\ref{eq:T_chi})
\begin{eqnarray}
    T_{(o)}{^r_t}&=&-K_{o} \nonumber \\
    &=&-f^2 T_{(o)}{^t_r} \nonumber \\
    T_{(o)}{^r_r}&=&\frac{K_{o}+Q_{o}}{f}+\frac{1}{2f}\int^r_{r_H}T_{(o)}{^\alpha_\alpha}(r')f'(r')dr' \nonumber \\
    &=&T_{(o)}{^\alpha_\alpha}-T_{(o)}{^t_t} \, . \label{eq:T_o}
\end{eqnarray}

To see how the gravitational anomaly should be compensated by some fluxes, we start with the effective action $W$ obtained by path integrating out the scalar field
\begin{align}
    W(g_{\mu\nu})=-i\log\left(\int \mathcal{D}\varphi e^{iS\left(\varphi,g_{\mu\nu}\right)}\right) \, .
\end{align}
\begin{svgraybox}
Since the fundamental theory is generally covariant, the variation of effective action $W$ under coordinate transformation $x^\mu\to x^\mu -\lambda^\mu$
\begin{eqnarray}
    -\delta W &=&\int d^2x \, \sqrt{-g_{(2)}}\lambda^\nu \nabla_\mu T^\mu_\nu \nonumber\\
    &=& 0 \, , \label{eq:variation_W}
\end{eqnarray}
implying that the anomaly in the near-horizon region (\ref{eq:anomaly_chiral}) has to be cancelled by some fluxes, and we will show that this gives the flux Hawking radiation. 
\end{svgraybox}

With the divergences of energy-momentum tensors (\ref{eq:divergence_near}) and (\ref{eq:divergence_outside}), the variation of the chiral effective action with neglecting the incoming modes is
\begin{eqnarray}
    -\delta W_{(\chi)+(o)} 
    &=& \int d^2x \, \Bigg\{\lambda^t \left[\partial_r\left(N^r_t H\right)+(T_{(o)}{^r_t}-T_{(\chi)}{^r_t}+N^r_t)\partial_r\Theta_+ \right] \nonumber \\
    &+& \lambda^r \left[\left(T_{(o)}{^r_r}-T_{(\chi)}{^r_r}\right)\partial_r\Theta_+\right] \Bigg\} \, . \label{eq:variation_W_chi_o}
\end{eqnarray}
It is noteworthy that the first total-derivative term $\partial_r\left(N^r_t H\right)$ is thought to be cancelled by the quantum effect of the incoming modes in the literature \cite{Iso:2006wa,Iso:2006xj,Murata:2006pt}, represented by \cite{Iso:2006ut}
\begin{eqnarray}
    T_{(L)}{^\mu_\nu}&=& -N^\mu_\nu   \, ,
\end{eqnarray}
from the corresponding gravitational anomaly with flipping the sign for incoming modes
\begin{eqnarray}
    \nabla_\mu T_{(L)}{^\mu_\nu}=-\frac{1}{\sqrt{-g_{(2)}}}\partial_\mu N^\mu_\nu \, .
\end{eqnarray}
By adding this incoming part, the divergence (\ref{eq:divergence_near}) becomes
\begin{eqnarray}
    \nabla_\mu\left(T_{(\chi)}{^\mu_\nu}H+T_{(L)}{^\mu_\nu}H\right) &=& \nabla_\mu\left(T_{(\chi)}{^\mu_\nu}+T_{(L)}{^\mu_\nu}\right)H+\left(T_{(\chi)}{^\mu_\nu}+T_{(L)}{^\mu_\nu}\right)\partial_\mu H \nonumber \\
    &=& -\left(T_{(\chi)}{^\mu_\nu}-N^\mu_\nu\right)\partial_\mu \Theta_+ \, ,
\end{eqnarray}
so the total derivative in (\ref{eq:variation_W_chi_o}) is canceled. With (\ref{eq:T_chi}), (\ref{eq:T_o}) and (\ref{eq:variation_W_chi_o}), the variation of the effective action is evaluated explicitly
\begin{eqnarray}
     -\delta W &= &\int d^2x \, \left[\lambda^t \left(-K_o+K_\chi+N^r_t\right)+\frac{\lambda^r}{f}\left(K_o+Q_o-K_\chi-Q_\chi\right) \right]\delta(r-r_H-\epsilon) \nonumber \\
     &=& 0 \, , \label{eq:delta_W_constraints}
\end{eqnarray}
for arbitrary $\lambda^\mu$, implying that their coefficients have to be zero, so we get two conditions
\begin{eqnarray}
    -K_o+K_\chi+N^r_t|_{r_H}  &=& 0 \nonumber \\
    K_o+Q_o-K_\chi-Q_\chi &=& 0 \, , 
\end{eqnarray}
which satisfy the relations
\begin{eqnarray}
    K_o&=&K_\chi+N^r_t|_{r_H}  \nonumber \\
    Q_o&=&Q_\chi-N^r_t|_{r_H}  \, . \label{eq:KQ_condition}
\end{eqnarray}
Such a shift of the energy-momentum tensor can be understood as a flux
\begin{eqnarray}
    \Phi&=&{\rm sign}(\eta)N^r_t|_{r_H} \nonumber \\
    &=&\frac{\kappa^2}{48\pi} \, . \label{eq:flux}
\end{eqnarray}

To finally fix the flux ($K_o$), we need to impose a vanishing condition for the covariant energy-momentum tensor at the horizon $\tilde{T}_{(\chi)}{^\mu_\nu}|_{r_H}=0$, and with (\ref{eq:Atr}), (\ref{eq:Atildetr}) and (\ref{eq:T_chi}) it is 
\begin{eqnarray}
    \tilde{T}_{(\chi)}{^r_t}&=& T_{(\chi)}{^r_t}+\int^r_{r_0} \tilde{A}_t(r')-A_t(r')dr' \nonumber \\
    &=& -K_{\chi}+\int^r_{r_H} A_t(r')dr'+\eta\frac{ff''-2f'^2}{192\pi} \, 
\end{eqnarray}
which $r_0$ is determined by $f'(r_0)=f''(r_0)=0$, and this gives $K_\chi=-\eta\frac{\kappa^2}{24\pi}$. Inserting this to (\ref{eq:KQ_condition}), we have
\begin{eqnarray}
    T_{(o)}{^r_t}&=&\eta\frac{\kappa^2}{48\pi} \nonumber \\
    &=& \eta \Phi \, . \label{eq:T^r_t_gravitaitonal_anomaly}
\end{eqnarray}
For an outgoing flux of real scalar field, we have $T_{(o)}{^r_t}=g_{tt}T_{(o)}{^{rt}}=g_{tt}\left(\partial_t\phi(t-r_*)\right)^2<0$, suggesting that $\eta=-1$ which is also consistent to the sign derived by the trace anomaly \cite{Christensen:1977jc} shown in section \ref{sec:trace_anomaly}, and the size of the flux outside the horizon is indeed $\Phi$. By comparing to the flux of thermal radiation with temperature $T$ in $(1+1)$ dimension: 
\begin{eqnarray}
    \Phi_{\rm Th}&=& \langle \rho v^r \rangle_{\rm Th} \nonumber \\
    &=& \int^{+\infty}_{0}\frac{dk}{2\pi} \frac{k}{e^{\frac{k}{T}}-1} \nonumber \\
    &=& \frac{\pi T^2}{12} \, ,
\end{eqnarray}
where $\langle \dots \rangle_{\rm Th}$ means taking a thermal average for the energy flux $\rho v^r$, we obtain the correct Hawking temperature $T_H=\frac{\kappa}{2\pi}$. It is noteworthy that this method only proves the flux as a thermal radiation, and more directly related thermal features such as spectrum cannot be shown explicitly \cite{Das:2007ru}. However, the gravitational-anomaly method can derive the thermal distribution in some cases such as higher-spin currents with a charged black hole \cite{Iso:2007hd,Iso:2007kt,Iso:2008sq}.

\begin{svgraybox}
    \subsubsection*{Remark on the original derivation by Robinson and Wilczek}
    There are a few differences between the original derivation in \cite{Robinson:2005pd} (see also \cite{Das:2007ru,Vagenas:2006qb}) and the one presented in this chapter, which is based on several subsequent papers \cite{Iso:2006wa,Iso:2006ut,Murata:2006pt}: 
    \begin{itemize}
        \item {\bf The discarded modes are outgoing modes instead of the ingoing ones}
        
        As the Boulware vacuum causes a diverging energy-momentum for a freely falling observer at the horizon, they chose to exclude outgoing ("horizon-skimming") modes attributing such a divergence, which also leads to a $(1+1)$-dimensional chiral theory. However, most of the subsequent papers adopt the exclusion of ingoing modes in \cite{Iso:2006wa}, and this difference was also discussed in the appendix of \cite{Iso:2006xj} by Iso, Morita and Umetsu, as well as an essay by Das, Robinson and Vagenas \cite{Das:2007ru}.

        As explained in \cite{Iso:2006xj}, both the outgoing and ingoing choices give the same answer in the Schwarzschild black hole, since the factors like $f'^2$ and $f''$ in (\ref{eq:N_components}) are unchanged. However, for the cases of charged or rotating black holes, excluding ingoing modes gives the correct Hawking radiation, whereas another choice cannot, so we adopt the former in this book.
        \\

        \item {\bf Consider also the region inside the horizon and the justification of the total-derivative term}
        
        \cite{Robinson:2005pd} also include the region inside the horizon, and $T^\mu_\nu$ (\ref{eq:Tmunu_separated}) is now separated into three parts
        \begin{eqnarray}
            T^\mu_\nu&=& \begin{cases}
            T_{(o)}{^\mu_\nu}\, , & r>r_H+\epsilon\\
            T_{(\chi)}{^\mu_\nu}\, , & |r-r_H|\leq \epsilon\\
            T_{(i)}{^\mu_\nu}\, , & r<r_H-\epsilon
            \end{cases} \nonumber \\
            &=&T_{(\chi)}{^\mu_\nu}H+ T_{(o)}{^\mu_\nu} \Theta_+ +T_{(i)}{^\mu_\nu}\Theta_- \, ,
        \end{eqnarray}
        where the Heaviside step functions $\Theta_\pm=\Theta(\pm r \mp r_H-\epsilon)$, and the boxcar (top-hat) function $H=1-\Theta_+-\Theta_-$. Correspondingly, the divergences (\ref{eq:divergence_near}) and (\ref{eq:divergence_outside}) now become
        \begin{eqnarray}
            \nabla_\mu \left(T_{(\chi)}{^\mu_\nu} H\right)&=& \nabla_\mu T_{(\chi)} {^\mu_\nu}H+T_{(\chi)}{^\mu_\nu}\partial_\mu H \nonumber \\
            &=&\partial_\mu\left(N^\mu_\nu H\right)-\left(T_{(\chi)}{^\mu_\nu}-N^{\mu}_\nu\right)\partial_\mu \left(\Theta_++\Theta_-\right) \, ,
        \end{eqnarray}
        and
        \begin{eqnarray}
            \nabla_\mu\left( T_{(o)}{^\mu_\nu}\Theta_+ + T_{(i)}{^\mu_\nu}\Theta_-\right) 
            &=& T_{(o)}{^\mu_\nu} \partial_\mu \Theta_+ +T_{(i)}{^\mu_\nu} \partial_\mu \Theta_- \, ,
        \end{eqnarray}
        respectively. The energy-momentum tensor in the region $|r-r_H|>\epsilon$ is conserved $\nabla_\mu T_{(o/i)}{^\mu_\nu}=0$, and this fixes the form of $T_{(o/i)}{^\mu_\nu}$:
        \begin{eqnarray}
            T_{(o/i)}{^r_t}&=&-K_{o/i} \nonumber \\
            &=&-f^2 T_{(o/i)}{^t_r} \nonumber \\
            T_{(o/i)}{^r_r}&=&\frac{K_{o/i}+Q_{o/i}}{f}+\frac{1}{2f}\int^r_{r_H}T_{(o/i)}{^\alpha_\alpha}(r')f'(r')dr' \nonumber \\
            &=&T_{(o/i)}{^\alpha_\alpha}-T_{(o/i)}{^t_t} \, . \label{eq:T_oi}
        \end{eqnarray}
        With these modifications, the variation of effective action (\ref{eq:delta_W_constraints}) now includes more terms involving $T_{(i)}{^\mu_\nu}$ and $\Theta_-$
        \begin{eqnarray}
            &&\delta W \nonumber \\
            &=& \int d^2x \, \Bigg\{\lambda^t \left[\partial_r\left(N^r_t H\right)+(T_{(o)}{^r_t}-T_{(\chi)}{^r_t}+N^r_t)\partial_r\Theta_+ + (T_{(i)}{^r_t}-T_{(\chi)}{^r_t}+N^r_t)\partial_r\Theta_-\right]\nonumber  \nonumber \\ 
            &+&\lambda^r \left[\left(T_{(o)}{^r_r}-T_{(\chi)}{^r_r}\right)\partial_r\Theta_++\left(T_{(i)}{^r_r}-T_{(\chi)}{^r_r}\right)\partial_r\Theta_-\right] \Bigg\} \, . \label{eq:delta_W_Robinson}
        \end{eqnarray}
        The total-derivative term $\partial_r\left(N^r_t H\right)$ in (\ref{eq:delta_W_Robinson}) is assumed to be vanishing when $\epsilon\to 0$ \cite{Robinson:2005pd,Robinson:2005ph}. But to the knowledge of the chapter's author, the explicit justification of this is unclear:
        \begin{eqnarray}
            \partial_r\left(N^r_t H\right)\approx 2\epsilon N^r_t \delta'(r-r_H) +\mathcal{O}(\epsilon^2)\, ,
        \end{eqnarray}
        as this also contributes a term with $\epsilon \delta'(r-r_H)$ which will be shown relevant to determine the flux. On the other hand, the same total-derivative term in (\ref{eq:variation_W_chi_o}) is argued to be cancelled by the quantum effect of the incoming modes \cite{Iso:2006wa}.
        \\ 

        \item {\bf More conditions to fix the flux}
        
        To fix the flux from (\ref{eq:delta_W_Robinson}), we need more conditions to determine $K_i$ and $Q_i$ in (\ref{eq:T_oi}), requiring the derivative of Heaviside step functions to be expanded to $\mathcal{O}(\epsilon)$
        \begin{eqnarray}
            \partial_\mu\Theta_{\pm}\approx\delta^r_\mu\left(\pm \delta(r-r_H)-\epsilon \delta'(r-r_H)\right)+\mathcal{O}(\epsilon^2)\, ,
        \end{eqnarray}
        and thus (\ref{eq:delta_W_Robinson}) (if the total-derivative term is neglected) becomes
        \begin{eqnarray}
            \nonumber \\
             \delta W &\approx &\int d^2x \, \Bigg\{\lambda^t\left[\left(K_o-K_i\right)\delta(r-r_H)-\epsilon\left(K_o+K_i-2K_\chi-2N^r_t\right)\delta'(r-r_H)\right] \nonumber \\
            &-&\lambda^r\Bigg[\frac{K_o+Q_o+K_i+Q_i-2K_\chi-2Q_\chi}{f}\delta(r-r_H)\nonumber \\
            &-&\epsilon\frac{K_o+Q_o-K_i-Q_i}{f}\delta'(r-r_H)\Bigg]\Bigg\} +\mathcal{O}(\epsilon^2) \, ,
        \end{eqnarray}
        By requiring $\delta W=0$ for arbitrary $\lambda^\mu$, we require all the coefficients of $\delta(r-r_H)$ and $\delta'(r-r_H)$ to be zero, giving four equations
        \begin{eqnarray}
            K_o-K_i &=& 0 \nonumber \\
            K_o+K_i-2K_\chi-2N^r_t|_{r_H} &=& 0 \nonumber \\
            K_o+Q_o+K_i+Q_i-2K_\chi-2Q_\chi &=& 0 \nonumber \\
            K_o+Q_o-K_i-Q_i &=& 0 \, ,
        \end{eqnarray}
        which satisfy the relations
        \begin{eqnarray}
            K_o&=&K_i=K_\chi+N^r_t|_{r_H}  \nonumber \\
            Q_o&=&Q_i=Q_\chi-N^r_t|_{r_H}  \, .
        \end{eqnarray}
        This gives the same condition to determine the Hawking radiation (\ref{eq:KQ_condition}), but more boundary condition, such as the vanishing ${\tilde{T}}{^\mu_\nu}$ at the horizon, is still needed to determine the value of $K_o$.
    \end{itemize}
\end{svgraybox}

It is noteworthy that the derivation with gravitational anomaly only requires the information near the horizon, such as the $(1+1)$-dimensional approximation (\ref{eq:1p1_effective}) and $N^\mu_\nu$ (\ref{eq:flux}) at the horizon. We will see in section \ref{sec:trace_anomaly} that the old derivation by trace anomaly requires more assumptions.

\subsection{Hawking radiation from the $(1+1)$-dimensional trace anomaly}\label{sec:trace_anomaly}
Before ending the section of anomaly method, we introduce an old derivation of Hawking radiation from the $(1+1)$-dimensional trace anomaly \cite{Christensen:1977jc}, and we will see some similarities and differences compared to gravitational-anomaly method in section \ref{sec:gravitaitonal_anomaly}.

Classically, the energy-momentum tensor of a $(1+1)$-dimensional massless scalar field
\begin{eqnarray}
    T^\mu_{\nu}=\partial^\mu \varphi \partial_\nu \varphi -\frac{\delta^\mu_{\nu}}{2}\partial^\alpha\phi\partial_\alpha\varphi \, ,
\end{eqnarray}
which clearly has zero trace $T^\alpha_\alpha=0$. However, when consider a (renormalized) quantum expectation value of the trace, this can acquire a non-zero value related to the background geometry \cite{Davies:1976ei}, as shown in (\ref{eq:trace_anomaly}). With this trace anomaly for the metric (\ref{eq:metric_rt}), Christensen and Fulling \cite{Christensen:1977jc} showed that the flux of the Hawking radiation is determined by the conservation
\begin{eqnarray}
    \nabla_\mu T^\mu_\nu =0  \, ,
\end{eqnarray}
which has been solved in (\ref{eq:T_o}) when we consider the outside region in the gravitational-anomaly method, and the solution depends on two constants $K$ and $Q$, as well as a function $T^\alpha_\alpha$. These three unknown constants and function are fixed now as follows, and there are some conditions different to the gravitational-anomaly method.
\begin{svgraybox}
\begin{itemize}
    \item The trace anomaly (\ref{eq:trace_anomaly}) fixes $T^{\alpha}_{\alpha}$ and thus the integral term in (\ref{eq:T_o})
    \begin{eqnarray}
        T{^\alpha_\alpha}&=&-\frac{f''}{24\pi} \nonumber \\
        \frac{1}{2f}\int^r_{r_H}T{^\alpha_\alpha}(r')f'(r')dr'&=&-\frac{f'^2-(2\kappa)^2}{96\pi f} \, .
    \end{eqnarray}
    It is noteworthy that the gravitational-anomaly method does not rely on $T^\alpha_\alpha$ to determine the integration constants in (\ref{eq:KQ_condition}), which is replaced by the gravitational anomaly in $\nabla_\mu T^\mu_\nu$.
    \item The $T_{uu}$ vanishes at least quadratically at the horizon
    \begin{eqnarray}
        |T_{uu}f^{-2}| |_{r_H}<\infty \, ,
    \end{eqnarray}
    for no singularity for a freely falling observer, implying that
    \begin{eqnarray}
        \frac{1}{4}\left(-fT^t_t+fT^r_r-2T^r_t\right)\Big|_{r_H}&=&K+\frac{Q}{2} \nonumber \\
        &=& 0 \, ,
    \end{eqnarray}
    where (\ref{eq:T_o}) is applied, and this gives one condition of $K$ and $Q$ (the definitions of $K$ and $Q$ in \cite{Christensen:1977jc} are different to ours: $K_{\rm there}=-K$ and $Q_{\rm there}=-2K-Q$). For the gravitational-anomaly method, this is replaced by the vanishing boundary condition for the covariant energy-momentum tensor at the horizon $\tilde{T}^r_t=0$.
    \item At spatial infinity, the energy density is equal to the flux $T^t_t=T^r_t$ because for the outgoing modes of scalar field
    \begin{eqnarray}
        T^t_t=T^r_t&=&- \left(\partial_t\phi(t-r)\right)^2 \, ,
    \end{eqnarray}
    which gives another condition of $K$ and $Q$:
    \begin{eqnarray}
        -K-Q-\frac{\kappa^2}{24\pi}=-K \, .
    \end{eqnarray}
    For the gravitational-anomaly method, it is not required to have such a matching at the infinity, which is replaced by a matching between the chiral and non-chiral parts at $r=r_H+\epsilon$ (near the horizon) in order to cancel the anomaly for the full theory. 
\end{itemize}
\end{svgraybox}
These conditions imply the same flux as the one by gravitational anomaly (\ref{eq:T^r_t_gravitaitonal_anomaly}), that
\begin{eqnarray}
    T^r_t&=& -K \nonumber \\
    &=&-\frac{\kappa^2}{48\pi} \, ,
\end{eqnarray}
and again this is consistent to the previous choice that $\eta=-1$ in (\ref{eq:N_mu_nu}).

The drawback of the trace anomaly is restricted to the case of massless scalar in the whole spacetime (not just the approximation near the horizon), and therefore this seems to be a special feature. This in fact motivates the derivation of gravitational anomaly, introduced in the previous section \ref{sec:gravitaitonal_anomaly}, which is applicable in more general cases. On the other hand, by comparing the consistent answers by the two anomaly methods, one can also justify some boundary conditions essential to derive the results, such as the vanishing covariant energy-momentum tensor at the horizon used in the gravitational anomaly \cite{Murata:2006pt}.

\section{The Green's function method}\label{sec:Green_fun_method}
Soon after Hawking's original derivation, the thermal feature of the Green's function for particles propagating in the black hole's spacetime was noticed \cite{Gibbons:1976es,Gibbons:1976pt}, and its mathematical properties were used in the well-known Hartle-Hawking path-integral derivation \cite{Hartle:1976tp}. The thermal character of Green's function has also been generalized to the cases of particle production in cosmology \cite{Chitre:1977ip,Gibbons:1977mu} and accelerated frame \cite{Troost:1977dw,Troost:1978yk}, suggesting the universality of this method in various flat or curved spacetimes. 

\subsection{Thermal Green's function}\label{sec:thermal_Green}
The thermal feature of Hawking radiation can be seen in the periodicity of Green's function when analytically continued to complex coordinates, consistent to the behavior of a thermal Greens's function \cite{Gibbons:1976es,Gibbons:1976pt}. We leave the detailed analysis of the analytical properties of Hawking radiation in section \ref{sec:path_integral_approach}, and here we first discuss the thermal case.

\begin{figure}
    \centering
    \includegraphics[width=0.6\textwidth]{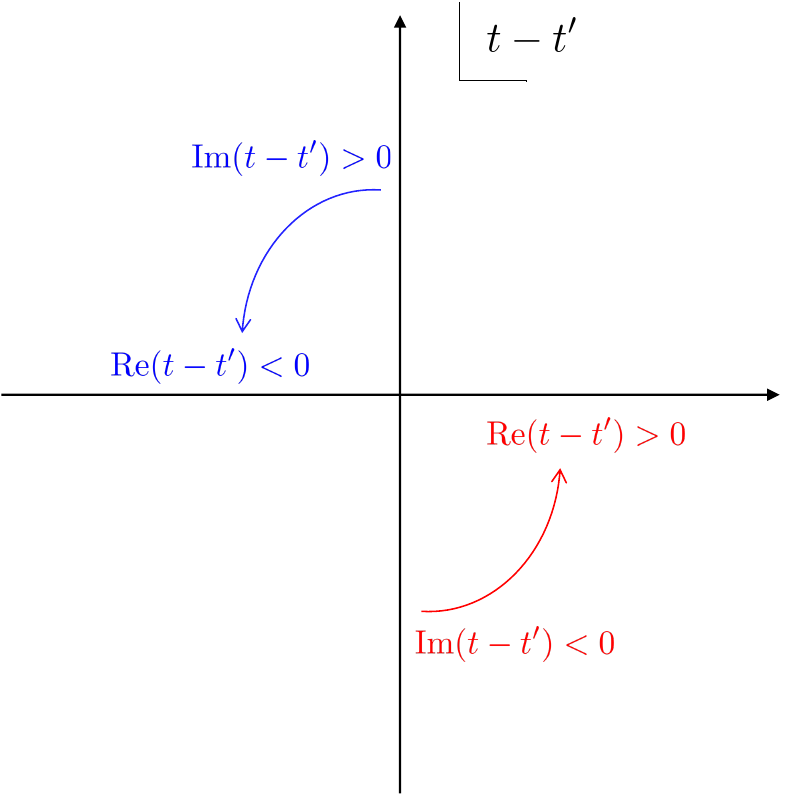}
    \caption{The time ordering defined in (\ref{eq:time-ordering}) agrees with the analytic continuation. \label{fig:time-ordering}}
\end{figure}

For a system in a thermal state described by the following density matrix
\begin{eqnarray}
    \rho=\frac{e^{-\beta H}}{{\rm Tr}\left(e^{-\beta H}\right)}\, ,
\end{eqnarray}
where $\beta=\frac{1}{T}$ is the inverse of temperature, and $H$ is the system's Hamiltonian, and the expectation value of operators is defined by
\begin{eqnarray}
    \langle \dots \rangle_\beta &=& {\rm Tr}(\rho \dots) \nonumber \\
    &=& \frac{{\rm Tr}\left(e^{-\beta H}\dots\right)}{{\rm Tr}\left(e^{-\beta H}\right)} \, .
\end{eqnarray}
In particular, for a scalar field in the Heisenberg picture
\begin{eqnarray}
    \phi(t,{\bf x})=e^{iHt}\phi(0,{\bf x})e^{-iHt}\, ,
\end{eqnarray}
the following two Green functions
\begin{eqnarray}
    G^+_T(t,{\bf x};t',{\bf x}')&=&i\frac{{\rm Tr}\left(e^{-\beta H}\phi(t,{\bf x})\phi(t',{\bf x}')\right)}{{\rm Tr}\left(e^{-\beta H}\right)}\nonumber \\
    &=&i\frac{{\rm Tr}\left(e^{-\beta H}\phi(t,{\bf x})e^{\beta H}e^{-\beta H}\phi(t',{\bf x}')\right)}{{\rm Tr}\left(e^{-\beta H}\right)} \nonumber \\
    &=& i\frac{{\rm Tr}\left(\phi(t+i\beta,{\bf x})e^{-\beta H}\phi(t',{\bf x}')\right)}{{\rm Tr}\left(e^{-\beta H}\right)} \nonumber \\
    &=& i\frac{{\rm Tr}\left(e^{-\beta H}\phi(t',{\bf x}')\phi(t+i\beta,{\bf x})\right)}{{\rm Tr}\left(e^{-\beta H}\right)}  \nonumber \\
    &=& G^-_T(t+i\beta,{\bf x};t',{\bf x}') \, , \label{eq:G_adv}
\end{eqnarray}
and similarly
\begin{eqnarray}
    G^-_T(t,{\bf x};t',{\bf x}')&=&i\frac{{\rm Tr}\left(e^{-\beta H}\phi(t',{\bf x}')\phi(t,{\bf x})\right)}{{\rm Tr}\left(e^{-\beta H}\right)}\nonumber \\
    &=& i\frac{{\rm Tr}\left(e^{-\beta H}e^{\beta H}\phi(t,{\bf x})e^{-\beta H}\phi(t',{\bf x}')\right)}{{\rm Tr}\left(e^{-\beta H}\right)} \nonumber \\
    &=& i\frac{{\rm Tr}\left(e^{-\beta H}\phi(t-i\beta,{\bf x})\phi(t',{\bf x}')\right)}{{\rm Tr}\left(e^{-\beta H}\right)}  \nonumber \\
    &=& G^+_T(t-i\beta,{\bf x};t',{\bf x}')\, , \label{eq:G_ret}
\end{eqnarray}
give the Kubo–Martin–Schwinger (KMS) condition for a system in thermal equilibrium \cite{Kubo:1957mj,Martin:1959jp}. From (\ref{eq:G_adv}) and (\ref{eq:G_ret}), the thermal Green function 
\begin{eqnarray}
    G_T(t,{\bf x};t',{\bf x}')&=&i\frac{{\rm Tr}\left(e^{-\beta H}T\phi(t,{\bf x})\phi(t',{\bf x}')\right)}{{\rm Tr}\left(e^{-\beta H}\right)} \, ,
\end{eqnarray}
has to be defined in the complex $t$-plane, and there is a subtlety in defining the time-ordering $T$ in a complex way \cite{Fulling:1987}:
\begin{eqnarray}
    T\phi(t,{\bf x})\phi(t',{\bf x}')=\begin{cases}
        \phi(t,{\bf x})\phi(t',{\bf x}') & {\rm Re}(t)>{\rm Re}(t'),\, {\rm Im}(t)={\rm Im }(t') \\
        \phi(t',{\bf x}')\phi(t,{\bf x}) & {\rm Re}(t)<{\rm Re}(t'),\, {\rm Im}(t)={\rm Im }(t') \\
        \phi(t,{\bf x})\phi(t',{\bf x}') & {\rm Im}(t)<{\rm Im }(t') \\
        \phi(t',{\bf x}')\phi(t,{\bf x}) & {\rm Im}(t)>{\rm Im }(t')  
    \end{cases}\label{eq:time-ordering}
\end{eqnarray}
where $|{\rm Im}(t'-t')|<\beta$. Such a definition puts smaller imaginary time to the left is consistent with the case to put larger real time to the left if ${\rm Im}(t)={\rm Im}(t')$, by the analytic continuation shown in figure \ref{fig:time-ordering}.
For ${\rm Im}(t-t')=-\epsilon$ (making ${\rm Im}(t-t'+i\beta)<\beta$), the thermal Green's function with the time ordering (\ref{eq:time-ordering}) is
\begin{eqnarray}
    G_T(t+i\beta,{\bf x};t',{\bf x}')&=&  G^-_T(t+i\beta,{\bf x};t',{\bf x}')\nonumber \\
    &=& G^+_T(t,{\bf x};t',{\bf x}') \nonumber \\
    &=& G_T(t,{\bf x};t',{\bf x}') \, ,
\end{eqnarray}
where (\ref{eq:G_adv}) is applied. Similarly, for ${\rm Im}(t-t')=\epsilon$ (making ${\rm Im}(t-t'-i\beta)>-\beta$),
\begin{eqnarray}
    G_T(t-i\beta,{\bf x};t',{\bf x}')&=&  G^+_T(t-i\beta,{\bf x};t',{\bf x}')\nonumber \\
    &=& G^-_T(t,{\bf x};t',{\bf x}') \nonumber \\
    &=& G_T(t,{\bf x};t',{\bf x}') \, ,
\end{eqnarray}
where (\ref{eq:G_ret}) is applied. Therefore, the thermal Green's function is periodic in the imaginary time with the period of inverse temperature $i\beta=i\frac{1}{T}$, and such a characteristic for a thermal system can also be seen in a system with a black hole.

\subsection{The path-integral approach}\label{sec:path_integral_approach}
We follow the path-integral approach by Hartle and Hawking \cite{Hartle:1976tp}, in which the Hawking radiation is manifested by comparing the amplitudes of emitting or absorbing a particle by a black hole. The feature of this approach is to obtain the propagator in the field-theory sense from the path integral of point particles via the Schwinger's proper time formalism \cite{Schwinger:1951nm,Schwartz:2014sze}, providing a more intuitive derivation compared to the direct approach with Green's function.

For a point particle, we choose its action as
\begin{eqnarray}
    S[x]=\frac{1}{4}\int^W_0 g_{\mu\nu}\frac{dx^\mu}{dw}\frac{dx^\nu}{dw}dw \, ,
\end{eqnarray}
where $w$ and $W$ are parameter times, which play the roles of the Schwinger's proper time in the curved spacetime. Such a quadratic form of action, which is different to the usual action by proper time, has advantages in calculating path integrals. The amplitude for a particle from $x'$ at $w=0$ to $x$ at $w=W$ is a path integral, obtained by integrals over spacetime in infinitely many small time intervals with size $\epsilon=\frac{W}{N+1}$:
\begin{eqnarray}
    & &F(W,x,x')\nonumber \\
    &=&\int \mathcal{D}x \, e^{iS[x]} \nonumber \\
    &=&\lim_{N\to \infty}\int \frac{d^4x_N}{A}\sqrt{-g(x_N)}\dots\int \frac{d^4x_1}{A}\sqrt{-g(x_1)} \exp\left(i\sum_{i=0}^N S\left(\epsilon,x_{i+1},x_i\right)\right) \, , \nonumber \\
    \label{eq:path_integral_F}
\end{eqnarray}
where $x_0=x'$ and $x_{N+1}=x$ correspond to the two end points, and the normalization is
\begin{eqnarray}
    A=\int d^4x \sqrt{-g(x)}e^{iS(\epsilon,x,x')}\, . \label{eq:normalization_A}
\end{eqnarray}

It is expected that the path integral (\ref{eq:path_integral_F}) is oscillating rapidly, and the standard procedure is to analytically continue the time variable to imaginary value such that the integral is well-defined. For a Schwarzschild black hole in the Kruskal coordinates
\begin{eqnarray}
    ds^2=\frac{32M^3e^{-\frac{r}{2M}}}{r}\left(-dT^2+dX^2\right)+r^2d\Omega^2 \, , \label{eq:Kruskal}
\end{eqnarray}
where the $T$ and $X$ satisfy 
\begin{eqnarray}
    T&=& \begin{cases}
        \left(\frac{r}{2M}-1\right)^{\frac{1}{2}} e^{\frac{r}{4M}}\sinh\left(\frac{t}{4M}\right) & r>2M \\
        \left(1-\frac{r}{2M}\right)^{\frac{1}{2}} e^{\frac{r}{4M}}\cosh\left(\frac{t}{4M}\right)  & r<2M \\
    \end{cases} \nonumber \\
    X&=& \begin{cases}
        \left(\frac{r}{2M}-1\right)^{\frac{1}{2}} e^{\frac{r}{4M}}\cosh\left(\frac{t}{4M}\right) & r>2M \\
        \left(1-\frac{r}{2M}\right)^{\frac{1}{2}} e^{\frac{r}{4M}}\sinh\left(\frac{t}{4M}\right)  & r<2M \\
    \end{cases} \nonumber \\
    -T^2+X^2&=&\left(\frac{r}{2M}-1\right)e^{\frac{r}{2M}} \, , \label{eq:XT_Kruskal}
\end{eqnarray}
the analytic continuation is done by letting $T=i\zeta$, and the parameter times $w$ and $W$ are continued to negative imaginary values $-i\omega$ and $-i\Omega$ respectively. The corresponding metric $\gamma_{\mu\nu}$ is now defined by
\begin{eqnarray}
    ds^2=\frac{32M^3e^{-\frac{r}{2M}}}{r}\left(d\zeta^2+dX^2\right)+r^2d\Omega^2\, , 
\end{eqnarray}
and another useful expression in the following calculation is the local Riemann normal coordinates around a point $x$ with a small deviation $z$
\begin{eqnarray}
    \gamma_{\mu\nu}\approx \delta_{\mu\nu}-\frac{1}{3}R_{\mu\alpha\nu\beta}z^\alpha z^\beta+\mathcal{O}(z^3) \, .
\end{eqnarray}
With these, the normalization factor $A$ (\ref{eq:normalization_A}) evaluated from $x$ to $x+z$ within a small interval $\epsilon$ is analytically continued to
\begin{eqnarray}
    A&=&\int d^4z \sqrt{\gamma(x+z)}e^{-S(\epsilon,x,x+z)}  \nonumber \\
    &\approx& \int d^4z \, \left(1-\frac{1}{6}R_{\mu\nu}z^\mu z^\nu+\mathcal{O}(z^3)\right)e^{-\frac{1}{4\epsilon}\delta_{\mu\nu}z^\mu z^\nu} \nonumber \\
    &\approx& 16\pi^2\epsilon^2 +\mathcal{O}(\epsilon^3) \, .
\end{eqnarray}
Instead of directly evaluating the path-integral (\ref{eq:path_integral_F}), solving the partial differential equation satisfied by it is an easier way, obtained by the change of the (analytically continued) amplitude $F(\Omega+\epsilon,x,x')$ with respect to the parameter time
\begin{eqnarray}
    F(\Omega+\epsilon,x,x')&=&\int \frac{d^4y}{A}\sqrt{\gamma(y)}e^{-S(\epsilon,x,y)} F(\Omega,y,x') \nonumber \\
    &=& \int \frac{d^4z}{A}\sqrt{\gamma(x+z)}e^{-S(\epsilon,x,x+z)} F(\Omega,x+z,x') \nonumber \\
    &\approx& \int \frac{d^4z}{A} e^{-\frac{1}{4\epsilon}\delta_{\mu\nu}z^\mu z^\nu} \Bigg[\left(1-\frac{1}{6}R_{\mu\nu}z^\mu z^\nu\right)\left(1+z^\alpha\partial_\alpha+\frac{1}{2}z^\alpha z^\beta\partial_\alpha\partial_\beta\right) \nonumber \\
    &+&\mathcal{O}(z^3)\Bigg]F(\Omega,x,x') \nonumber \\
    &=&F(\Omega,x,x')+\epsilon \delta^{\mu\nu}\left(\partial_\mu\partial_\nu-\frac{R_{\mu\nu}}{3}\right)F(\Omega,x,x')\, , \label{eq:F_change_normal}
\end{eqnarray}
where $\mathcal{O}(z^3)$ terms are neglected because they contribute higher orders in $\epsilon$ after integrating over $z$. (\ref{eq:F_change_normal}) obtained with the local normal coordinates can be generalized to the covariant form by replacing $\delta^{\mu\nu}\to \gamma^{\mu\nu}$ and $\partial_\mu\partial_\nu\to\nabla_\mu\nabla_\nu$, and thus we have the partial differential equation
\begin{eqnarray}
    \frac{\partial F}{\partial \Omega}&=&\left(\gamma^{\mu\nu}\nabla_\mu\nabla_\nu-\frac{R}{3}\right)F \nonumber \\
    &=&\gamma^{\mu\nu}\nabla_\mu\nabla_\nu F  \, ,
\end{eqnarray}
as $R=0$ for the Schwarzschild black hole. By analytically continuing back to the original metric (\ref{eq:Kruskal}), we have
\begin{eqnarray}
    i\frac{\partial F}{\partial W}
    &=&-g^{\mu\nu}\nabla_\mu\nabla_\nu F  \, , \label{eq:PDE_F}
\end{eqnarray}
which has the form of the Schr\"odinger equation and satisfies the initial condition
\begin{eqnarray}
    F(0,x,x')=\frac{1}{\sqrt{-g}}\delta^4(x-x')  \, . \label{eq:initial_condition_F}
\end{eqnarray}
With a small $W$, $F(W,x,x')$ has a WKB ansatz
\begin{eqnarray}
    F(W,x,x')&\approx& N(W,x,x') e^{iS(W,x,x')} \nonumber \\
    &=& N(W,x,x') e^{i\frac{s(x,x')}{4W}} \, ,
\end{eqnarray}
where $s(x,x')$ is the square of distance between $x$ and $x'$. For flat space, one can easily check that 
\begin{eqnarray}
    N(W)=\frac{i}{(4\pi W)^2} \, ,
\end{eqnarray}
gives an exact solution of (\ref{eq:PDE_F}). For general curved spacetime, initial condition (\ref{eq:initial_condition_F}) implies that the solution for small $W$ is dominated by the neighborhood of $x$, which can be approximated by local flatness, so it is expected that
\begin{eqnarray}
    N(W,x,x')\approx \frac{D(x,x')}{W^2} \, , \label{eq:small_W_N}
\end{eqnarray}
generally, and a more detailed discussion can be found in \cite{Morette:1951zz}.
We follow the Schwinger's proper time formalism to obtain the Feynman propagator between the two points $x$ and $x'$ by a weighted sum of the amplitude over all possible $W$
\begin{eqnarray}
    G_F(x,x')=\lim_{\epsilon\to 0^+}i\int^{+\infty}_0 dW \, e^{-im^2W-\frac{\epsilon}{W}} F(W,x,x') \, , \label{eq:propagator}
\end{eqnarray} 
where $m$ is the particle's mass, and we introduce a small suppression factor $e^{-\frac{\epsilon}{W}}$ to regulate the integral for avoiding the divergence at $W\to 0$ due to (\ref{eq:small_W_N}). With (\ref{eq:PDE_F}), one can easily check that
\begin{eqnarray}
    \left(g^{\mu\nu}\nabla_\mu\nabla_\nu-m^2\right)G_F&=& i\int^{+\infty}_0 dW \, e^{-im^2W} \left(g^{\mu\nu}\nabla_\mu\nabla_\nu-m^2\right)F(W,x,x') \nonumber \\
    &=&\int^{+\infty}_0 dW \, e^{-im^2W} \left(\frac{\partial}{\partial W}-im^2\right)F \nonumber \\
    &=& \int^{+\infty}_0 dW \frac{\partial}{\partial W}\left(e^{-im^2W}F\right) \nonumber \\
    &=& -F(0,x,x') \nonumber \\
    &=& -\frac{1}{\sqrt{-g}}\delta^4(x-x') \, ,
\end{eqnarray}
where the initial condition (\ref{eq:initial_condition_F}) and boundary condition $\lim_{W\to+\infty}F(W,x,x')=0$ are applied.\footnote{It relies on the fact that the parabolic equation (\ref{eq:PDE_F}) spreads the initial localized data (\ref{eq:initial_condition_F}) at late time (large $W$) as $F(W,x,x')\approx \mathcal{O}(W^{-2})$.} Therefore, the propagator $G_F(x,x')$ defined in (\ref{eq:propagator}) is indeed the Green's function of the Klein-Gordon equation.

To derive Hawking radiation, the relationship between the amplitudes of emitting and absorbing a particle by a black hole is needed, and this is determined by the analytical properties of the propagator $G_F(x,x')$ as follows:
\begin{svgraybox}
    \begin{itemize}
        \item $G_F(x,x')$ is periodic in ${\rm Im}(t)$ with a period $8\pi M$, which can be easily seen from the periodicity of $X$ and $T$ in the Kruskal coordinates (\ref{eq:XT_Kruskal}), consistent to the thermal Green's function discussed in section \ref{sec:thermal_Green} with temperature $T_H=\frac{1}{8\pi M}$.
        \item For $x$ a point with fixed $(r,\theta,\varphi)$ inside the horizon and $x'$ a fixed point outside the black hole, $G_F(x,x')$ is analytical in a strip in the complex $t$ plane with $-4\pi M\leq {\rm Im}(t)\leq 0$. By deforming the integration contour for calculating the amplitude of emitting a particle to ${\rm Im}(t)=-4\pi M$, the amplitude of absorbing is obtained.
    \end{itemize}
\end{svgraybox}

\begin{figure}
    \centering
    \includegraphics[width=0.9\textwidth]{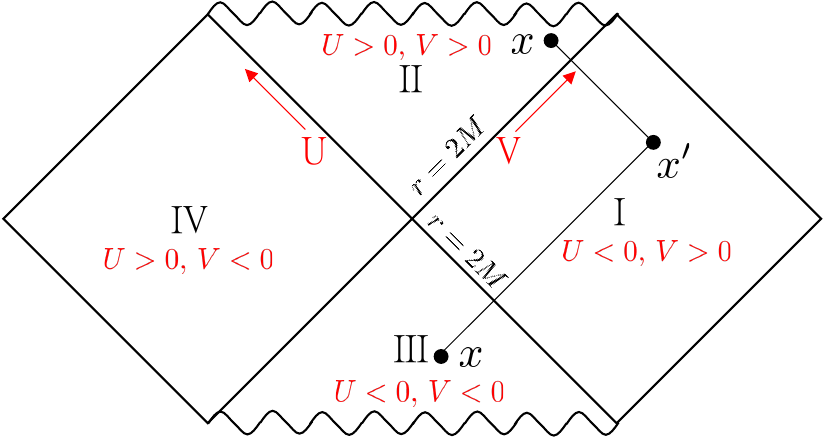}
    \caption{The Penrose diagram for a Schwarzschild black hole. We denote $x'$ as an observer in region I, and $x$ is the point of emitting (region II) or absorbing (region III) particles.\label{fig:BH_penrose_diagram}}
\end{figure}

Now we demonstrate the second point, starting by proving the analyticity. From (\ref{eq:propagator}), we know that the singular points of $G_F(x,x')$ are contributed by the small-$W$ part of the integrand, where the regularization with $\epsilon$ is important, and we have seen that this part has the WKB form with summing over all geodesics connecting $x$ and $x'$
\begin{eqnarray}
    F(W,x,x')\approx \sum_c\frac{D_c(x,x')}{W^2}e^{i\frac{s_c(x,x')}{4W}} \, .
\end{eqnarray}
With this, we can calculate $G_F(x,x')$ with $m=0$ by separating the integral by a singular part $[0,W_0]$ ($W_0$ is small) and an analytical part $[W_0,+\infty]$ denoted by $G_{F,0}$:
\begin{eqnarray}
    G_F(x,x')\approx -i \sum_c \frac{e^{i\frac{s_c(x,x')}{4W_0}}}{s_c(x,x')+i\epsilon}D_c(x,x')+G_{F,0} \, ,
\end{eqnarray} 
so the singular points are located at $s_c(x,x')+i\epsilon=0$. To see the locations of these points for null geodesics from $x'$ to $x$, it is convenient to define the null variables
\begin{eqnarray}
    U&=&T-X \nonumber \\
    V&=&T+X \nonumber \\
    ds^2&=&-32\frac{M^3}{r}e^{-\frac{r}{2M}}dUdV+r^2d\Omega^2 \, , \label{eq:ds2_UV}
\end{eqnarray} 
and four regions are defined according to the signs of the $(U,V)$ pair, according to the Penrose diagram in figure \ref{fig:BH_penrose_diagram}.
Therefore, a null geodesic with $s(x,x')=0$ connecting regions I and II is $V=V_0$. Now with the $i\epsilon$ prescription, the singular points are determined by\footnote{Do not be confused with the notation that $s(x,x')$ is the square of distance.}
\begin{eqnarray}
    s(x,x')&\approx& \frac{\partial s}{\partial V}(V-V_0)+\mathcal{O}(\epsilon^2) \nonumber \\
    &=&-i\epsilon \, , \label{eq:s_epsilon_V}
\end{eqnarray}
where for the simplest case with $d\Omega^2=0$ in (\ref{eq:ds2_UV})\footnote{For more general cases with $\Delta\varphi\neq 0$ is considered in \cite{Hartle:1976tp}.}
\begin{eqnarray}
    \frac{\partial s}{\partial V}&=& -\int^{U(x)}_{U(x')} \frac{32M^3}{r}e^{-\frac{r}{2M}} dU \nonumber \\
    &<& 0 \, ,
\end{eqnarray}
where we use the fact that $U(x)-U(x')>0$ since
\begin{eqnarray}
    U=\begin{cases}
        -\left(\frac{r}{2M}-1\right)^{\frac{1}{2}}e^{\frac{r-t}{4M}}<0 & {\rm region\ I \ }(r>2M)\\
        \left(1-\frac{r}{2M}\right)^{\frac{1}{2}}e^{\frac{r-t}{4M}}>0 & {\rm region\ II \ }(r<2M) \\
        \left(1-\frac{r}{2M}\right)^{\frac{1}{2}}e^{\frac{r-t}{4M}}<0 & {\rm region\ III \ }(r<2M) \, , \label{eq:U_regions}
    \end{cases}
\end{eqnarray}
and therefore we have ${\rm Im}(V-V_0)>0$ from (\ref{eq:s_epsilon_V}). From the relationship between $V$ and $(r,t)$
\begin{eqnarray}
    V=\begin{cases}
        \left(\frac{r}{2M}-1\right)^{\frac{1}{2}}e^{\frac{r+t}{4M}}>0 & {\rm region\ I \ }(r>2M)\\
        \left(1-\frac{r}{2M}\right)^{\frac{1}{2}}e^{\frac{r+t}{4M}}>0 & {\rm region\ II \ }(r<2M) \\
        \left(1-\frac{r}{2M}\right)^{\frac{1}{2}}e^{\frac{r+t}{4M}}<0 & {\rm region\ III \ }(r<2M)  \, , \label{eq:V_regions}
    \end{cases} 
\end{eqnarray}
we know that the singular points correspond to ${\rm Im}(t)=\epsilon$ in the region II.
On the other hand, for a point $x$ in the region III with ${\rm Im}(t)=-4\pi M$, the corresponding null geodesic with $s(x,x')=0$ is $U=U_0$. Similar to (\ref{eq:s_epsilon_V}), the singular points in $U$ is determined by
\begin{eqnarray}
    s(x,x')&\approx&\frac{\partial s}{\partial U}(U-U_0)+\mathcal{O}(\epsilon^2)\nonumber \\
    &=&-i\epsilon \, ,
\end{eqnarray} 
and 
\begin{eqnarray}
    \frac{\partial s}{\partial U}&=&-\int^{V(x)}_{V(x')} \frac{32M^3}{r}e^{-\frac{r}{2M}} dV \nonumber \\
    &>&0 \, ,
\end{eqnarray}
where $V(x)-V(x')<0$ from (\ref{eq:V_regions}), so we have ${\rm Im}(U-U_0)<0$. From (\ref{eq:U_regions}), we know that the corresponding singular points in the region III has ${\rm Im}(t)=-4\pi M-\epsilon$. Therefore, the two cases with the regions II and III prove the fact that $G_F(x,x')$ is analytical in a strip with $-4\pi M\leq {\rm Im}(t)\leq 0$.

With the analyticity of $G_F$, we can calculate the amplitude of emitting and absorbing particles from the black hole. As a demonstration of the relationship between the amplitude and the propagator, we pick a field $\phi$ to calculate. Consider initial and final states, $|f_i\rangle$ and $|g_f\rangle$ respectively \cite{Gibbons:1976pt}
\begin{eqnarray}
    |f_i\rangle &=&\int d\Sigma^\mu f(x)\overleftrightarrow{{\partial}_\mu}\phi(x)|0_-\rangle\nonumber \\
    |g_f\rangle &=&\int d{\Sigma'}^{\nu'} g(x')\overleftrightarrow{{\partial}_{\nu'}}\phi(x')|0_+\rangle \, ,
\end{eqnarray}
where $|0_\pm\rangle$, $\Sigma$ and $\Sigma'$ are the vacua and hypersurfaces at initial and final times, and $f\overleftrightarrow{{\partial}_\mu} g= f (\partial_\mu g) - (\partial_\mu f) g $. Therefore, the amplitude from the initial to final states corresponds to a detector at $x'$ receiving a particle from $x$
\begin{eqnarray}
    i\frac{\langle g_f|f_i\rangle}{\langle 0_+|0_-\rangle}&=& \int \int \frac{d\Sigma^\mu d{\Sigma'}^{\nu'}}{\langle 0_+|0_-\rangle} \, \langle 0_+|\Bigg[ g^*\partial_{\nu'}\left(f\phi(x')\partial_\mu \phi(x)-\phi(x')\phi(x)\partial_\mu f\right)\nonumber \\ 
    &-& \left(f\phi(x')\partial_\mu\phi(x)-\phi(x')\phi(x)\partial_\mu f\right)\partial_{\nu'}g^*\Bigg] |0_-\rangle \nonumber \\
    &=&\int \int d\Sigma^\mu d{\Sigma'}^{\nu'} \, g^*_{\nu'}\left(f\partial_\mu G_F-G_F \partial_\mu f\right)-\left(f\partial_\mu G_F-G_F \partial_\mu f\right)\partial_{\nu'}g^* \nonumber \\
    &=& \int \int d\Sigma^\mu d{\Sigma'}^{\nu'} \, g^* \overleftrightarrow{{\partial}_{\nu'}}\left(f\overleftrightarrow{\partial_\mu} G_F\right) \nonumber \\
    &=& -\int \int d\Sigma^\mu d{\Sigma'}^{\nu'} \,  g^*(x') \overleftrightarrow{{\partial}_{\nu'}}\left(G_F(x',x)\overleftrightarrow{\partial_\mu} f(x)\right)  \, , \label{eq:amplitude_integral}
\end{eqnarray}
where we used
\begin{eqnarray}
    G_F(x',x)=i\frac{\langle 0_+|T\phi(x')\phi(x)|0_-\rangle}{\langle 0_+|0_-\rangle}\, .
\end{eqnarray}
(\ref{eq:amplitude_integral}) implies that we can determine the amplitude of emission by the propagator that obtained previously in (\ref{eq:propagator}).
Due to the time translational symmetry of the Schwarzschild black hole, we expect that there exists modes with $g(x')\propto e^{-iE t'}$ and $f(x)\propto e^{-iE t}$, together with the form of the propagator
\begin{eqnarray}
    G_F(x',x)=G_F(0,{\bf r}';t-t',{\bf r})\ .
\end{eqnarray}
By choosing the hypersurface with constant $r$ in region II, the amplitude (\ref{eq:amplitude_integral}) is proportional to
\begin{eqnarray}
    \mathcal{E}_E({\bf r}',{\bf r})&=&\int^{+\infty}_{-\infty}dt\, e^{-iEt}G_F(0,{\bf r}';t,{\bf r})  \nonumber \\
    &=&e^{-4\pi M E}\int^{+\infty}_{-\infty}dt\, e^{-iEt}G_F(0,{\bf r}';t-4i\pi M,{\bf r}) \nonumber \\
    &=&e^{-4\pi M E} \tilde{\mathcal{E}}_E({\bf r}',{\bf r})  \, ,  \label{eq:amplitude_deform}
\end{eqnarray}
where we deform the contour to ${\rm Im}(t)=-4\pi M$ in the second line, and $\tilde{\mathcal{E}}_E({\bf r}',{\bf r})$ corresponds to the contribution to the amplitude from a point $x$ in the region III to $x'$ in the region I. As the region III is the time reversal of the region II, we can also interpret $\tilde{\mathcal{E}}_E({\bf r}',{\bf r})$ contributing the amplitude from $x'$ in the region I to the reflected point of $x$ in the region II (going into the black hole), obtained by changing back to ${\rm Im}(t)=0$. After taking the absolute square of the amplitude, from (\ref{eq:amplitude_deform}) we know that the probabilities of emitting and absorbing particles are related to
\begin{eqnarray}
    P_{\rm emit}&=&e^{-8\pi M E}P_{\rm absorb} \nonumber \\
    &=&e^{-\frac{E}{T_H}}P_{\rm absorb} \, , \label{eq:P_emit_absorb_path_integral}
\end{eqnarray}
giving the Boltzmann factor with the Hawking temperature $T_H=\frac{1}{8\pi M}$.

\section{Discussion and conclusion}\label{sec:conclusion}
In this chapter, we have presented three types of derivations of Hawking radiation: the tunneling, anomaly and Green's function methods. Different setups in each type have also been compared, such as the null geodesic and Hamilton-Jacobi in the tunneling type, and the gravitational and trace anomalies in the anomalous type. The relationship between different types of methods can be seen as follows:
\begin{svgraybox}
    \begin{itemize}
        \item {\bf Tunneling vs Green's function (path integral)}
        
        It is easy to see that the tunneling method is the WKB approximation of the Hartle-Hawking path-integral approach \cite{Vanzo:2011wq}, and the physical quantity to calculate is essentially the same: the probability of emitting particles in (\ref{eq:Gamma_S_B_H}) and (\ref{eq:P_emit_absorb_path_integral}). Some authors also consider that the path-integral approach is also as semi-classical as the tunneling method in the sense that the wave modes of the quantum field are not required, and one of the differences is that the former requires the Kruskal extension in (\ref{eq:amplitude_deform}), whereas in the latter it is not necessary \cite{Shankaranarayanan:2000qv,Srinivasan:1998ty} (note that a complex path with the Kruskal extension can also be used, as demonstrated in figure \ref{fig:complex_path}). Another similarity (and also difference) of the two methods is the use of analytical continuation: deforming the path to evaluate the singular integral in the tunneling method (\ref{eq:iepsilon}) and analytically continuing the Green function to obtain the ratio of emitting to absorbing probability of radiation in (\ref{eq:amplitude_deform}).
        \\
        
        \item {\bf Tunneling vs gravitational anomaly}
        
        Clearly, the physical quantities calculated in these two methods are different: tunneling focuses on probability, whereas anomaly relies on the flux of radiation to deduce the Hawking temperature. However, it has been suggested that the two methods can be connected via chirality \cite{Banerjee:2008sn,Banerjee:2009wb}. The idea is to argue that the Kruskal variables $(T,X)$ inside and outside the black holes (\ref{eq:XT_Kruskal}) are related via
        \begin{eqnarray}
            t_{\rm in}&=&t_{\rm out}-2i\pi M \nonumber \\
            r_{*,{\rm in}}&=&r_{*,{\rm out}}+2i\pi M \nonumber \\
            u_{\rm in}&=& u_{\rm out}-4i\pi M \nonumber \\
            v_{\rm in}&=& v_{\rm out} \, , \label{eq:matching_uv}
        \end{eqnarray}
        where the relationship $e^{\frac{r_*}{4M}}=\left(\frac{r}{2M}-1\right)^\frac{1}{2}e^{\frac{r}{4M}}$ is used. Therefore, the probability for a left-handed (ingoing) mode moves inward is
        \begin{eqnarray}
            P_{L}&=&\left|e^{-i\omega v_{\rm in}}\right|^2 \nonumber \\
            &=&\left|e^{-i\omega v_{\rm out}}\right|^2 \nonumber \\
            &=& 1 \, ,
        \end{eqnarray}
        whereas for a right-handed (outgoing) mode tunneling to outside is 
        \begin{eqnarray}
            P_{R}&=& \left|e^{-i\omega u_{\rm in}}\right|^2 \nonumber \\
            &=& \left|e^{-i\omega \left(u_{\rm out}-4i\pi M\right)}\right|^2 \nonumber \\
            &=& e^{-\frac{\omega}{T_H}} \, ,
        \end{eqnarray}
        so the tunneling probability can be derived from the correct chiral mode. However, the connection between this probability and the tunneling calculation with classical action in section \ref{sec:tunneling} is not manifest with matching the variables in (\ref{eq:matching_uv}).
        On the other hand, there are some situations, such as the Rindler spacetime for an accelerating observer, that the tunneling and anomaly methods give different results \cite{Akhmedova:2008au}, suggesting that the relationship between the two methods is still subtle.
    \end{itemize}
\end{svgraybox}

From the various derivations of Hawking radiation, many features of black holes reflect their rich physical and mathematical structures, and we should expect that there are still many unexplored aspects. The Hawking radiation is thus a useful effect to verify the consistency of different points of view on black holes, a subtle intersection between gravity and quantum physics.

\begin{acknowledgement}
CM Sou thanks the helpful discussion with Ali Akil and Wei-Xiang Feng, and the support from the Shuimu Tsinghua scholar program, NSFC under Grant No. 12275146, the National Key R\&D Program of China (2021YFC2203100), and the Dushi Program of Tsinghua University.
\end{acknowledgement}


\end{document}